\documentclass[journal]{IEEEtran}

\usepackage{cite}
\usepackage{array}
\usepackage{color}
\usepackage{float}
\usepackage[cmex10]{amsmath}
\usepackage{nomencl}						
\usepackage[normalem]{ulem}			        
\usepackage{diagbox}						
\usepackage{slashbox}						
\usepackage{colortbl}						
\usepackage{multirow}						
\usepackage{tabularx}
\usepackage{placeins}
\usepackage{algorithm}
\usepackage{mathrsfs}
\usepackage{mathdots}
\usepackage{amssymb}
\usepackage{amsthm}
\usepackage{arydshln}
\usepackage{color}
\usepackage[noend]{algpseudocode}
\usepackage{bm}
\usepackage{url}
\usepackage[hidelinks]{hyperref}
\usepackage[T1]{fontenc}
\newcommand{\subparagraph}{}
\usepackage{fancyhdr} 
\usepackage{extarrows}
\usepackage{enumitem}
\usepackage{booktabs}
\usepackage{threeparttable}
\usepackage{ragged2e}

\ifCLASSINFOpdf
  \usepackage[pdftex]{graphicx}
	\graphicspath{{./figure/}}
  \DeclareGraphicsExtensions{.pdf,.jpeg,.png}
\else
  \usepackage[dvips]{graphicx}
  \graphicspath{{./figure/}}
  \DeclareGraphicsExtensions{.eps}
\fi

\ifCLASSOPTIONcompsoc
  \usepackage[caption=false,font=normalsize,labelfont=sf,textfont=sf]{subfig}
\else
  \usepackage[caption=false,font=footnotesize]{subfig}
\fi

\newcommand{\figref}[1]{\figurename~\ref{#1}}
\newcommand{\tabref}[1]{Table~\ref{#1}}

\ifCLASSOPTIONonecolumn

\else

\fi

\begin{document}
	\clearpage
\thispagestyle{plain}
\twocolumn[
\begin{@twocolumnfalse}
	\begin{center}
		\vspace{4cm}
		\LARGE
		\textbf{Copyright Statements}
		\vspace{1cm}
	\end{center}
	\justifying
	\Large
	This work has been submitted to the IEEE for possible publication. Copyright may be transferred without notice, after which this version may no longer be accessible.
	\justifying
\end{@twocolumnfalse}
]
	
	\clearpage
\bstctlcite{IEEEexample:BSTcontrol}
\setcounter{page}{1}

\title{Impedance-based Root-cause Analysis: Comparative Study of Impedance Models and 
Calculation of Eigenvalue Sensitivity}
\author{Yue~Zhu, \IEEEmembership{Student Member, IEEE}, Yunjie~Gu, \IEEEmembership{Senior Member, IEEE}, Yitong~Li, \IEEEmembership{Member, IEEE}, Timothy~C.~Green, \IEEEmembership{Fellow, IEEE}}

\IEEEaftertitletext{\vspace{-1.5\baselineskip}}

\ifCLASSOPTIONpeerreview
	\maketitle 
\else
	\maketitle
\fi

\thispagestyle{fancy}
\chead{This work has been submitted to the IEEE for possible publication. Copyright may be transferred without notice, after which this version may no longer be accessible.}
\rhead{\thepage}
\cfoot{}
\renewcommand{\headrulewidth}{0pt}
\pagestyle{fancy}

\begin{abstract}
Impedance models of power systems are useful when state-space models of apparatus such as inverter-based resources (IBRs) have not been made available and instead only black-box impedance models are available. For tracing the root causes of poor damping and tuning modes of the system, the sensitivity of the modes to components and parameters are needed. The so-called critical admittance-eigenvalue sensitivity based on nodal admittance model has provided a partial solution but omits meaningful directional information. The alternative whole-system impedance model yields participation factors of shunt-connected apparatus with directional information that allows separate tuning for damping and frequency, yet do not cover series-connected components. This paper formalises the relationships between the two forms of impedance models and between the two forms of root-cause analysis. The calculation of system eigenvalue sensitivity in impedance models is further developed, which fills the gaps of previous research and establishes a complete theory of impedance-based root-cause analysis. The theoretical relationships and the tuning of parameters have been illustrated with a three-node passive network, a modified IEEE 14-bus network and a modified NETS-NYPS 68-bus network, showing that tools can be developed for tuning of IBR-rich power systems where only black-box impedance models are available.
\end{abstract}


\makenomenclature
\nomenclature[01]{$Y^{\text{nodal}}, Y^{\text{nodal}}_{ki}$}{nodal admittance model and its elements}
\nomenclature[02]{$Z^{\text{sys}}, Z^{\text{sys}}_{ki}$}{whole-system impedance model and its elements}
\nomenclature[03]{$Y_{\text{N}}$}{network nodal admittance matrix}
\nomenclature[04]{$A,a_{mn}$}{state-space matrix and its elements}
\nomenclature[05]{$\lambda$}{eigenvalue of the state-space matrix $A$}
\nomenclature[06]{$\gamma$}{critical admittance-eigenvalue, the zero-eigenvalue of $Y^{\text{nodal}}(\lambda)$}
\nomenclature[07]{$S_{\lambda},S_{\lambda,ki}$}{eigenvalue sensitivity matrix and its elements}
\nomenclature[08]{$S_{\gamma},S_{\gamma,ki}$}{critical admittance-eigenvalue sensitivity matrix and its elements}
\nomenclature[09]{$w_{\gamma}, u_{\gamma}$}{left and right eigenvectors of $Y^{\text{nodal}}(\lambda)$ corresponding to $\gamma$, normalized as $ w_{\gamma}^\top u_{\gamma}=1$.}
\nomenclature[10]{$Y^{\text{nodal}}_{\det}(\lambda)$}{$\det( Y^{\text{nodal}}(s)) \mid_{s=\lambda}$, the determinant of $Y^{\text{nodal}}$ at $\lambda$}
\nomenclature[11]{$Y^{\text{nodal}}_{\det}{}'(\lambda)$}{derivative of $Y^{\text{nodal}}_{\det}(s)$ to $s$ at $s=\lambda$.}
\nomenclature[12]{$\text{Res}_\lambda Z^{\text{sys}}$}{residue of $Z^{\text{sys}}$ at $\lambda$}
\nomenclature[13]{$s_{\lambda,y}$}{admittance sensitivity factor}
\nomenclature[14]{$s_{\lambda,\rho}$}{parameter sensitivity factor}
\nomenclature[15]{$\text{eig}(\cdot)$}{the eigenvalues of the matrix}
\nomenclature[16]{$\text{tr}(\cdot)$}{trace of the matrix}
\nomenclature[17]{$\text{adj}(\cdot)$}{adjugate of the matrix}
\nomenclature[18]{$\langle \cdot,\cdot \rangle$}{Frobenius inner product}
\nomenclature[19]{$^\top$}{transpose}
\nomenclature[20]{$^*$}{conjugate transpose: $(\cdot)^* = \overline{({\cdot})}^\top$}


\printnomenclature[1.9cm]


\section{Introduction}
State-space models, which are white-box models, have been the mainstay of small-signal analysis of power systems for assessing stability (through eigenvalues and pole-zero maps)\cite{Singh2013task}, for root-cause analysis (through eigenvalue sensitivity and participation factors)\cite{Perez1982Selective, Smed1993Feasible} and for damping design (through pole-placement, linear matrix inequalities etc)\cite{Pal2000Linear}. State-space methods can be applied to any physical system and have been applied to large conventional power systems and networks dominated by inverter based resources (IBRs) such as microgrids\cite{Bottrell2013Dynamic}. However, the state-space method can be difficult to apply in practice if IBRs are present because the differential equations describing the controllers of IBRs are not generally openly available due to commercial confidentiality of the control implementation. Further, generic models of IBRs are hard to apply because there are a variety of approaches to IBR control available, because IBR features do not necessarily scale well with power ratings and because of the lack of agreed approaches to model reduction given the interacting dynamics of subsystems in the IBR. Manufacturers do release black-box models such as compiled models for EMT simulation \cite{AECOM2017EMT} and sometimes also impedance spectrum models $Z(j\omega)$, that relate perturbations of current to changes of voltage at the terminals\cite{Sun2019Modeling, Gu2020Motion,Li2021Impedance,Wu2021Development}. These impedance spectra are transfer functions between the two variables at the electrical terminal and they capture the dynamics relevant to interactions with the grid.

Impedance models have been popular in the field of power electronics for analysis of interactions between source and load power converters \cite{Wen2014Smallsignal}. They normally take the form of an output impedance of a source, $Z_S\left(j\omega\right)$ and an input admittance of a load, $Y_L\left(j\omega\right)$, which combine to give a closed-loop transfer function in the form  $1/(I+Z_S{\left(j\omega\right)Y}_L\left(j\omega\right))$ to which Nyquist stability criterion can be applied. Originally this was applied to low-voltage DC distribution but has been extended to 3-phase AC systems modelled in the $d\text{-}q$ frame. However, the source-load partition cannot readily be extended to large, meshed networks. Alternative formulations of impedance (or admittance) models of meshed networks have been created and fall into two categories: nodal-loop model\cite{Ebrahimzadeh2018Participation,Zhan2019Modal}  in which circuit equations from either nodal or loop analysis are assembled in a matrix format, and whole-system model\cite{ZhangImpedance2020,Gu2021Impedance, Zhu2021Participation} in which network branch admitances and impedances of apparatus at nodes are combined in a feedback loop and all elements contain dynamics of the whole-system. These two formats will be discussed fully in section \ref{section_networked}. For power systems, such models comprise matrices of impedances representing the whole network where each entry is either a transfer-function or a frequency spectrum, referred  to a global reference frame. 

When considering large power systems through impedance models, in nodal-loop model and whole-system model formats, it is desirable to replace Nyquist criteria and phase or gain margins with eigenvalue analysis and pole-zero assessment supplemented with modal analysis such as sensitivity and participation analysis. Creating such analysis for impedance models starts to build relationships between impedance models and state-space models, and starts to look inside the black-box impedance models and achieve almost the same transparency as white-box state-space models. This has been described as a grey-box approach\cite{Zhu2021Participation}. However, several important issues deserve further attention:
\begin{itemize}[leftmargin=*]
    \item[1)] The relationship between nodal-loop model and whole-system model and their underlying association with oscillatory modes should be compared and clarified, so their relative merits can be established.
    \item[2)] For nodal-loop model, a method described as sensitivity of the critical admittance-eigenvalue has been put forward for tracing root-causes of poor damping. However, as will be discussed in section \ref{critical admittance-eigenvalue section}, this method involves the loss of some important directional information that hampers identification of appropriate tuning actions. For whole-system model, the grey-box approach is thus far only applicable for analysing participation of shunt-connected devices. Extension to identify participation of series-connected apparatus in branches is needed to be able to tune a static series synchronous compensator (SSSC) or the series element of a unified power flow controller (UPFC).
    \item[3)] The original eigenvalue sensitivity, i.e. the sensitivity of state-space eigenvalue describing the modes, has not been solved for impedance models in any of the previous research.
\end{itemize}

This paper sets out to address the above issues. It will present the full analytical relationship between root-cause analysis methods in two forms of impedance model and propose a unified approach to calculate the eigenvalue sensitivity for root-cause analysis of instability. The paper is organised as follows. Nodal-loop model and whole-system model are discussed and compared in Section II. In Section III, the critical admittance-eigenvalue sensitivity is reviewed with its limitations discussed. From this, it will be shown that a further term can be identified to relate critical admittance-eigenvalue sensitivity to other forms of eigenvalue sensitivity analysis and create a unified view of root-cause analysis. In Section IV, the theories are verified through numerical case-studies of a 3-node passive circuit, a modified IEEE 14-bus network and a modified NETS-NYPS 68-bus network. The last section concludes the paper.

\begin{figure*}
	\centering
	{\includegraphics[scale = 0.48]{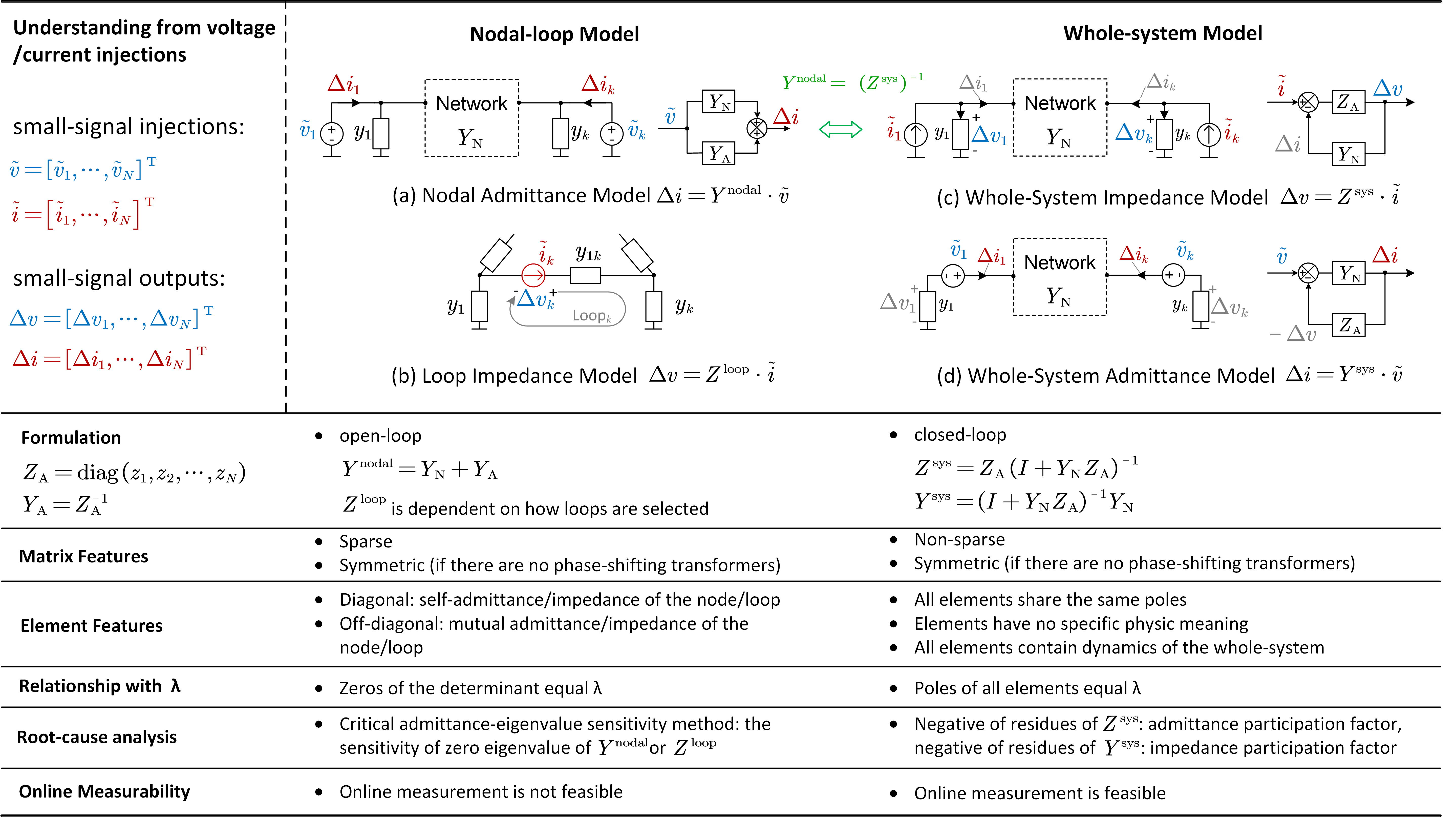}}
	\caption{The four types of networked impedance models and their characteristics.}
	\label{fig1_formation}	
\end{figure*}

\section{Review of the Networked Impedance Models} \label{section_networked}
Recent research work has extended the applicability of impedance spectrum models to meshed network through creation of matrices in which each entry in a matrix is a transfer function or a frequency spectrum aligned to a global synchronous frame. Currently, there are two major categories of such networked impedance modes: nodal-loop model and whole-system model. An overview is given in \figref{fig1_formation} of the formulation, the main characteristics, and the relationships between nodal-loop model and whole-system model.

\subsection{Nodal-loop Model}
Nodal-loop model uses the well-established nodal or loop forms of circuit analysis to generate a set of circuit equations assembled in matrix form. We refer to such models as a nodal admittance model $Y^{\text{nodal}}$ and a loop impedance model $Z^{\text{loop}}$.

$Y^{\text{nodal}}$ is of the same form as the nodal admittance matrix $Y_{\text{N}}$ widely used for power flow calculations in power systems except that (i) instead of each entry in $Y_{\text{N}}$ being a single value evaluated at the operating frequency, the entries in $Y^{\text{nodal}}$ are all spectra or in the form of small-signal transfer functions at an operating point of the power system, (ii) instead of treating each item of shunt-connected apparatus as a zero-impedance voltage source (consequently seen as an open-circuit in $Y_{\text{N}}$), $Y^{\text{nodal}}$ treats each apparatus as an admittance entry and incorporates them with other parts of the network. As shown in \figref{fig1_formation}(a), $Y^{\text{nodal}}$ expresses how a vector of voltage perturbations $\Tilde{v}$ applied at the nodes of the grid creates a vector of corresponding changes in nodal currents $\Delta i$. Such formation is a simple open-loop relationship as
\begin{equation}
\label{check_y_def}
    Y^{\text{nodal}}=Y_{\text{N}}+Y_{\text{A}}
\end{equation}

Similarly, $Z^{\text{loop}}$ expresses how a vector of current perturbations applied to each loop current of the grid creates a vector of corresponding additional loop voltage drops around the loop, as shown in \figref{fig1_formation}(b). Although broadly a dual of the nodal admittance model, the loop impedance model can be formulated in a variety of ways for a given grid and without a systematic approach to the choice of loop currents. It is difficult to manipulate this model in further analysis.

Both $Y^{\text{nodal}}$ and $Z^{\text{loop}}$ will express the resonances or modes in the system, in other words, they will reflect the eigenvalue of the state-space model. It has been proved \cite{Semlyen1999sDomain} that the zeros of the determinant of $Y^{\text{nodal}}$ or $Z^{\text{loop}}$ are equal elements of the vector of state-space eigenvalues $\lambda$. For the sake of brevity, we also use $\lambda$ to represent a single eigenvalue of the state-space model, and define $\det(Y^{\text{nodal}}(s)) \triangleq Y^{\text{nodal}}_{\det}(s)$ as a transfer function formed by the determinant, such that
\begin{equation}
    \label{det_zero}
    Y^{\text{nodal}}_{\det}(\lambda) = 0.
\end{equation}
One benefit of using the nodal-loop model is that each entry in the matrix is mapped to an aggregation of some specific components in the network, so offers explicit geographic information. For example, for $Y^{\text{nodal}}$, the diagonal entry $Y^{\text{nodal}}_{ii}$ represent all admittances terminating at node $i$, whereas the off-diagonal entry $Y^{\text{nodal}}_{ki}$ is the negative of the sum of admittances between node $k$ and node $i$. By studying eigenvalue sensitivity with respect to entries of $Y^{\text{nodal}}$, the root-cause of the oscillation can be easily located whether the root-cause is a shunt or series-connected item. Recognising this, a sensitivity method in nodal-loop model was proposed \cite{Wilsun2005Critical} and named critical mode sensitivity. The method seeks to replicate the analysis of the state-space matrix $A$, and investigates the sensitivity of the 'zero' eigenvalue of $Y^{\text{nodal}}$ with respect to an element $Y^{\text{nodal}}_{ki}$. The related issues will be discussed further in section \ref{critical admittance-eigenvalue section}.

Another important observation is that perturbation of a nodal voltage of $Y^{\text{nodal}}$ can be considered as a theoretical step but as a practical exercise is problematic. In practice, nodes will have stiff voltage sources present, hence perturbing the node voltage with a perturbation in parallel to conduct an online measurement of $Y^{\text{nodal}}$ is not feasible. Similarly, measuring $Z^{\text{loop}}$ online with a current perturbation in series is not feasible.

\subsection{Whole-system model}

The formulation of a model by identification of a source impedance and load admittance was feasible for simple power supply systems but difficult to apply in a meshed grid with intermingled sources and loads so other formulations have been sought. A useful separation is between, on the one hand, the shunt-connected apparatus appearing at nodes (which includes generators, such as synchronous machines and inverters, and loads) and on the other hand the lines and cables of the branches of the network that connect nodes \cite{Gu2021Impedance}. Based on this separation, the whole-system impedance model $Z^{\text{sys}}$ and the whole-system admittance model $Y^{\text{sys}}$ can be formed. Taking $Z^{\text{sys}}$ as an example, the system is first separated into a diagonal impedance matrix $Z_{\text{A}}$, where the diagonal-entries are the impedances of apparatus at each node, and the network nodal admittance matrix $Y_{\text{N}}$ containing the admittances of branches. The model is then formulated with a virtual nodal injection of current $\tilde{i}$ which causes a change in the apparatus voltage $\Delta v$, which in turn creates a feedback effect by causing a change of current flowing into the network $\Delta i$. This feedback arrangement is illustrated in \figref{fig1_formation}(c), and the response $\Delta v$ is
\begin{equation}
\label{eq_Delta_v}
    \Delta v=Z_{\text{A}}(I+Y_{\text{N}} Z_{\text{A}})^{-1}\cdot \tilde{i},
\end{equation}
where $I$ is identity matrix. (\ref{eq_Delta_v}) gives rise to the definition of $Z^{\text{sys}}$:
\begin{equation}
\label{eq_hat_Z_form}
    Z^{\text{sys}}=Z_{\text{A}}(I+Y_{\text{N}} Z_{\text{A}})^{-1}.
\end{equation}
Similarly, the formulation of whole-system admittance model is shown in \figref{fig1_formation}(d) and defined as 
\begin{equation}
    Y^{\text{sys}}= (I+Y_{\text{N}} Z_{\text{A}})^{-1} Y_{\text{N}}.
\end{equation}

The elements of $Y^{\text{sys}}$ and $Z^{\text{sys}}$ are all transfer functions which have a common set of poles and those poles are also identical to the poles of $\left(sI-A\right)^{-1}$ from the state-space model. As each element contains the information of the dynamic characteristics of the whole system, such model is referred as whole-system model. It is further reported in \cite{Zhu2021Participation} that the residues of the diagonal elements $Y^{\text{sys}}_{kk}$ 
equal the impedance participation factors of the apparatus connected in series with the $k$-th voltage injection source, and that the residues of $Z^{\text{sys}}_{kk}$ lead to the admittance participation factors of apparatus connected in parallel with the $k$-th current injection source. Compared with sensitivity method in nodal-loop model, the participation analysis in whole-system model creates a fuller view of how changes in parameters affect the damping and natural frequency of a mode. The details of the difference mentioned here will be explored in Section \ref{SectionII_sensitivity}. Note that that the participation analysis for whole-system model in \cite{Zhu2021Participation} did not extend to series-connected components.

From \figref{fig1_formation}, it can be seen by changing the voltage injections into current injections, $Y^{\text{nodal}}$ will become $Z^{\text{sys}}$, hence we have
\begin{equation}
\label{ZY_inverse}
    (Z^{\text{sys}})^{-1} = Y^{\text{nodal}}.
\end{equation}
The above relationship can also be easily proved mathematically from \eqref{check_y_def} and \eqref{eq_hat_Z_form}:
\begin{equation}
    (Z^{\text{sys}})^{-1} = (I+Y_{\text{N}} Z_{\text{A}}) Z_{\text{A}}^{-1}
    = Y_{\text{A}} + Y_{\text{N}} = Y^{\text{nodal}},
\end{equation}
noting that $Y_{\text{A}}=Z_{\text{A}}^{-1}$ is a diagonal matrix of apparatus admittance. Note that (\ref{ZY_inverse}) does not hold when $s=\lambda$ because  $\lambda$ is a singularity of $Y^{\text{nodal}}$. It is also worth remarking that there is no evidence of a general relationship between $Z^{\text{loop}}$ and $Y^{\text{sys}}$.

In contrast to the difficulty of measuring $Y^{\text{nodal}}$ and $Z^{\text{loop}}$ online with parallel voltage or series current perturbations, measuring $Z^{\text{sys}}$ and $Y^{\text{sys}}$ online with parallel current or series voltage perturbations is feasible.

Comparing across all models in \figref{fig1_formation}, one can conclude that for systematic analysis, nodal analysis is preferred because there is an exclusive definition for a given power system, which is not the case for loop analysis because loops can be selected in several different ways. For measurement of a model, it is preferable to connect a current injection source in parallel with node-connected apparatus since series-connection of a voltage perturbation source requires breaking into the original structure and inserting a transformer in series to facilitate the injection. For the further examinations, the models $Y^{\text{nodal}}$ and $Z^{\text{sys}}$ will be the main focus. 

\section{Impedance-based Root-cause analysis}
\label{SectionII_sensitivity}

Previous research \cite{Wilsun2005Critical,Huang2007Modal,LI2018RapidModal, Ebrahimzadeh2018Participation,Zhan2019Modal} discussed the concept and applications of critical admittance-eigenvalue sensitivity in $Y^{\text{nodal}}$ and $Z^{\text{loop}}$ with the aim of providing similar information to eigenvalue sensitivity. However, several issues related to critical admittance-eigenvalues have not been fully addressed. This section first reviews the critical admittance-eigenvalue sensitivity method and discusses its limitations. The formulation of eigenvalue sensitivity through impedance models in a way that is fully equivalent to that in state-space models is then set-out, and compared with the results of critical admittance-eigenvalue sensitivity. The relationship between eigenvalue sensitivity and the residues of $Z^{\text{sys}}$ is also revealed, which provides straightforward access to eigenvalue sensitivity with respect to any components' admittance. The grey-box approach of \cite{Zhu2021Participation} is extended to the sensitivity analysis to facilitate tracing of root-causes to different depths, i.e., apparatus and parameters, culminating in a complete theoretical approach to tracing root-causes in impedance models.

\subsection{Review of critical admittance-eigenvalue sensitivity $\frac{\partial\gamma}{\partial Y^{\text{nodal}}_{ki}}$}
\label{critical admittance-eigenvalue section}
The concept of a critical admittance-eigenvalue, also referred to in the literature as a critical resonance mode, a critical eigenvalue, or a critical mode, was first given in \cite{Wilsun2005Critical} where it was defined as the smallest eigenvalue of $Y_{\text{N}}$ at the resonance frequency of the mode of interest. Work in \cite{Ebrahimzadeh2018Participation, Zhan2019Modal} extended the scope of the critical admittance-eigenvalue to $Y^{\text{nodal}}$, and defined as the zero-eigenvalue of $Y^{\text{nodal}}$ at $s=\lambda$. Because such an `eigenvalue' is an eigenvalue of $Y^{\text{nodal}}$, we draw a careful distinction between it and $\lambda$, and choose the term 'critical admittance-eigenvalue' to refer to it. In this paper, we use $\lambda$ to represent the eigenvalues of state-space matrix $A$, referred to simply as eigenvalues, and use $\gamma$ to represent the critical admittance-eigenvalues. Here we also set out  definitions of $\lambda$ and $\gamma$ to help clarify the concepts.

An oscillatory mode $\lambda$ of the system is defined as:
\begin{equation}
    \lambda \in \text{eig}(A), \lambda=\sigma \pm j\omega,
\end{equation}
where $\sigma$ refers to the damping and $\omega$ refers to the natural frequency of the mode. For each $\lambda$, there is a corresponding critical admittance-eigenvalue $\gamma$ which is an eigenvalue of $Y^{\text{nodal}}(\lambda)$ defined as:
\begin{equation}
    \gamma \in \text{eig}\left( Y^{\text{nodal}}(\lambda) \right), \gamma =0.
\end{equation}
It is certain that a zero-valued critical admittance-eigenvalue exists because $\lambda$ is a zero of the determinant of $Y^{\text{nodal}}$, i.e., $Y^{\text{nodal}}_{\det}(\lambda)=0$ and $Y^{\text{nodal}}(\lambda)$ is not full-rank, thus there is at least one eigenvalue of $Y^{\text{nodal}}(\lambda)$ that is equal zero. Here we also introduce a very important premise: $\lambda$ is assumed to be a non-repeated eigenvalue of $A$. Under this premise, the rank of $Y^{\text{nodal}}(\lambda)$ is $N-1$, hence there is one and only one $\gamma$ corresponding to each $\lambda$. Such a premise was implicitly applied in the previous literature but not specifically mentioned.

Previous research stated that $\gamma$ is the main factor determining the characteristics of the mode being examined, but the proof of this finding has some flaws. Most previous research proves it by using the idea of modal current injection at the modal frequency. However $Y^{\text{nodal}}$ is a singular matrix at $s=\lambda$ and therefore its inverse matrix does not exist, neither does the inverse of its diagonalised matrix, a fact overlooked in the previous proofs. To avoid this difficulty, we prove this conclusion using a new method based on small-signal perturbation, as shown in Appendix \ref{critcial_proof}. The proof establishes that for a variation of a physical parameter in the system, the corresponding variation $|\Delta \gamma|$ is proportional to the variation of the mode $|\Delta \lambda|$. Consequently, the sensitivity $\frac{\partial\gamma}{\partial Y^{\text{nodal}}_{ki}}$ also reflects the sensitivity of the mode with respect to $Y^{\text{nodal}}_{ki}$. By extending such sensitivity to the whole network, a critical admittance-eigenvalue sensitivity matrix is then defined as $S_{\gamma}$, in which the entry in the $i$-th row and the $k$-th column, $S_{\gamma,ik}$, is the sensitivity of $\gamma$ with respect to the $(k,i)$ element of $Y^{\text{nodal}}(\lambda)$, as shown below:
\begin{equation}
\label{lambda_c}
S_{\gamma,ik}=\frac{\partial \gamma}{\partial Y^{\text{nodal}}_{ki}},
\end{equation}
It was proven in \cite{Huang2007Modal} that the matrix $S_{\gamma}$ can be directly calculated as the outer product of $ u_{\gamma}$ and $ w_{\gamma}$ (the left and right eigenvectors of $Y^{\text{nodal}}(\lambda)$ corresponding to $\gamma$) as:
\begin{equation}
\label{tensor}
S_{\gamma}= u_{\gamma}\otimes w_{\gamma}= u_{\gamma} w_{\gamma}^\top.
\end{equation}
This finding is encouraging because it has the same format as the sensitivity analysis in a state-space model: the outer product of left and right vectors \cite{kundur1994power}. Besides this, a major benefit of $S_{\gamma}$ is that the entry of $Y^{\text{nodal}}$ maps to specific physical components in the system, as mentioned in Section \ref{section_networked}. By comparing $\frac{\partial\gamma}{\partial Y^{\text{nodal}}_{ki}}$, system operators can determine which components play a dominant role in $\gamma$, thus providing intuition as to the root-cause of the oscillatory mode $\lambda$.

However, several key issues related to the critical admittance-eigenvalue sensitivity have not been addressed:
\begin{itemize}[leftmargin=*]
    \item[1)] The eigenvalue of interest, $\lambda$, is mostly likely a complex conjugate pair, i.e., $\lambda=\sigma \pm j\omega$. However, $\frac{\partial \gamma}{\partial Y^{\text{nodal}}_{ki}}$ cannot indicate how a component admittance affects $\sigma$ and $\omega$ as separate parts of $\lambda$, and so does not provide fully useful information on how to tune the parameter to shift $\lambda$ in the desired direction on the complex plane.
	\item[2)] The sensitivity values in $S_{\gamma}$ are complex but it has not been established how to interpret the real and imaginary parts and previous studies tend to resort to comparing the absolute values. It is also not clear how to compare the sensitivity in a three-phase system, where the sensitivity values are $2\times2$ matrix blocks in $d\text{-}q$ frame.
	\item[3)] $Y^{\text{nodal}}$ cannot be measured online, hence equation (\ref{tensor}) can not be used in a measurement-based situation.
\end{itemize}

\subsection{Calculation of the eigenvalue sensitivity $\frac{\partial \lambda}{\partial Y^{\text{nodal}}_{ki}}$}\label{section2.2}
To strictly evaluate how impedance of network components affect $\lambda$, the eigenvalue sensitivity $\frac{\partial \lambda}{\partial Y_{ki}}$ should be found. In a similar fashion to $S_\gamma$, we define the eigenvalue sensitivity matrix as $S_{\lambda}$, in which the element in the $j$-th row $k$-th column is
\begin{equation}
\label{Slabmda_def}
S_{\lambda,ik}=\frac{\partial \lambda}{\partial Y^{\text{nodal}}_{ki}}.
\end{equation}
Now we determine the value of $\frac{\partial \lambda}{\partial Y^{\text{nodal}}_{ki}}$.
By using the method of eigenvalue perturbation on (\ref{det_zero}), it can be proved that
\begin{equation}
\label{Sensitivity_matrix}
S_{\lambda}=-\frac{\mathrm{adj(}Y^{\text{nodal}}(\lambda ))}{Y^{\text{nodal}}_{\det}{}'\left( \lambda \right)}.
\end{equation}
The proof of (\ref{Sensitivity_matrix}) is given in Appendix \ref{proof-1} where $Y^{\text{nodal}}_{\det}{}'$ is the derivative of $Y^{\text{nodal}}_{\det}(s)$ at $s=\lambda$ defined in (\ref{Y_det_prime}). Equation (\ref{Sensitivity_matrix}) provides a direct method to calculate eigenvalue sensitivity from $Y^{\text{nodal}}$.

Now We seek to establish the formal relationship between $S_\lambda$ and $S_\gamma$. Because $\lambda$ is non-repeated, the rank of $Y^{\text{nodal}}(\lambda)$ is $N-1$, thus the rank of its adjunct matrix $\text{adj}( Y^{\text{nodal}}(\lambda))$ is $1$. It is known that a rank-1 matrix can be expressed as the outer product of two vectors. Hence we have
\begin{equation}
	\label{adj_Y_xy}
	\text{adj}( Y^{\text{nodal}}(\lambda))=x\otimes y=xy^\top,
\end{equation}
where $x$ and $y$ are two column-vectors of dimension $N$. Equation (\ref{adj_Y_xy}) has the same format as (\ref{tensor}). According to the proof in Appendix \ref{proof-2},
\begin{equation}
	\label{adj_Y}
	\text{adj}\left( Y^{\text{nodal}}\left( \lambda \right) \right) =xy^\top=\text{tr}\left( \text{adj}\left( Y^{\text{nodal}}\left( \lambda \right) \right) \right) \cdot  u_{\gamma}  w_{\gamma}^\top.
\end{equation}
Combining (\ref{lambda_c})-(\ref{adj_Y}) yields
\begin{equation}
	\label{Relationship}
	\begin{split}
		S_{\lambda}&=\xi \cdot S_{\gamma}=\xi  u_{\gamma} w_{\gamma}^\top
		\\
		\frac{\partial \lambda}{\partial Y^{\text{nodal}}_{ki}}&=\xi \cdot \frac{\partial \gamma}{\partial Y^{\text{nodal}}_{ki}}
	\end{split}
\end{equation}
where $\xi$ is a coefficient with a value of
\begin{equation}
\label{xi_cal}
\xi =-\frac{\text{tr}\left( \text{adj}\left( Y^{\text{nodal}}\left( \lambda \right) \right) \right)}{Y^{\text{nodal}}_{\det}{}'\left( \lambda \right)}.
\end{equation}

Equations (\ref{Relationship}) and (\ref{xi_cal}) reveal the relationship between critical admittance-eigenvalue sensitivity $\frac{\partial\gamma}{\partial Y^{\text{nodal}}_{ki}}$ and eigenvalue sensitivity $\frac{\partial\lambda}{\partial Y^{\text{nodal}}_{ki}}$: the two values are differ by a coefficient $\xi$. Since the mode of interest, $\lambda$, is usually a conjugate complex pair, it is clear from (17) that $\xi$ will accordingly be a conjugate complex pair, which contains directional information. By omitting $\xi$, the term $\frac{\partial \gamma}{\partial Y^{\text{nodal}}_{ki}}$ studied in the previous literature loses any meaning in terms of its direction in complex plane and hence cannot reveal a components' influence on the $\sigma$ and $\omega$ parts of $\lambda$.
As a result, components with relatively large magnitude of critical admittance-eigenvalue sensitivity may possibly, but not necessarily, affect the $\lambda$-mode. Comparing with the grey-box approach proposed in \cite{Zhu2021Participation}, critical admittance-eigenvalue sensitivity is equivalent to layer-1 of the grey-box. In contrast, the eigenvalue sensitivity $\frac{\partial \lambda}{\partial Y^{\text{nodal}}_{ki}}$ can precisely show how component admittance affects the mode in both damping and natural frequency and is comparable to layer-2 and layer-3 of the grey-box method.

\subsection{Complete theory of impedance-based root-cause analysis}
To compute the value of $\frac{\partial \lambda}{\partial Y^{\text{nodal}}_{ki}}$, it is straightforward to apply (\ref{Sensitivity_matrix}) or (\ref{Relationship}). However, as mentioned before, the elements in $Y^{\text{nodal}}$ cannot be measured readily online because small-signal voltage perturbations cannot be injected in parallel with stiff voltage sources. This is an important factor limiting the practical application of eigenvalue sensitivity in impedance models. On the other hand, $Z^{\text{sys}}$ can be measured online, and the residues of the diagonal-elements in $Z^{\text{sys}}$ can lead to the admittance participation factor of shunt-connected components. This observation lead us to explore the relationship between eigenvalue sensitivity and the residues of $Z^{\text{sys}}$. 

Since $\lambda$ is a non-repeated pole of $Z^{\text{sys}}$, according to the definitions, the residues of $Z^{\text{sys}}$ at $s=\lambda$ can be expressed as 
\begin{equation}
\mathrm{Res}_{\lambda}Z^{\mathrm{sys}}=\lim_{s\rightarrow \lambda} (s-\lambda )Z^{\mathrm{sys}}(s)
\end{equation}
Substituting (\ref{ZY_inverse}) yields
\begin{equation}
\label{res_Zsys_mid}
    \begin{split}
        \mathrm{Res}_{\lambda}Z^{\mathrm{sys}}&=\lim_{s\rightarrow \lambda} \left( (s-\lambda )Y^{\mathrm{nodal}}(s)^{-1} \right)
\\
&=\lim_{s\rightarrow \lambda} \left( \frac{s-\lambda}{Y_{\det}(s)}\mathrm{adj}\left( Y^{\mathrm{nodal}}(s) \right) \right) 
\\
    \end{split}
\end{equation}
Combined with the fact in (\ref{det_zero}), L'Hôpital's rule can be applied on (\ref{res_Zsys_mid}) and yields
\begin{equation}
\label{residue}
\mathrm{Res}_{\lambda}Z^{\text{sys}}=\frac{\mathrm{adj}\left( Y^{\text{nodal}}\left( \lambda \right) \right)}{Y^{\text{nodal}}_{\det}{}'\left( \lambda \right)},
\end{equation}
The detailed proof of equation (\ref{residue}) can be found in Appendix B in \cite{Zhu2021Participation}. Combining (\ref{Sensitivity_matrix}) and (\ref{residue}), it is clear to have
\begin{equation}
	\label{Res2Sens}
	S_{\lambda}=-\mathrm{Res}_{\lambda}Z^{\text{sys}}.
\end{equation}
The finding in equation (\ref{Res2Sens}) indicates a practical route for determining the sensitivity, because the spectra of the entries in $Z^{\text{sys}}$ can be measured online, and the poles and residues can be identified from the spectra by vector fitting techniques \cite{Gustavsen1999Rational}. This method also provides useful flexibility for comparing sensitivities where only some specific elements in $S_\lambda$ need to be compared. A system operator could choose to partially measure $Z^{\text{sys}}$, that is, the relevant elements of the matrix only, rather than acquiring the full matrix of the networked impedance model. Such flexibility cannot be achieved in equations (\ref{Sensitivity_matrix}) or (\ref{Relationship}).

Based on (\ref{Res2Sens}), eigenvalue sensitivity analysis can be implemented for the whole system and provide access to the state-space eigenvalue sensitivities with respect to admittance of any component in the system. For a component with admittance $y$, the eigenvalue sensitivity with respect to $y$ can be derived as
\begin{equation}
\label{shunt-series1}
\frac{\partial \lambda}{\partial y}=\sum_{k,i}^N{\left( \frac{\partial \lambda}{\partial Y^{\text{nodal}}_{ki}}\frac{\partial Y^{\text{nodal}}_{ki}}{\partial y} \right)}=-\mathrm{tr}\left( \mathrm{Res}_{\lambda}Z^{\text{sys}}\cdot \frac{\partial Y^{\text{nodal}}}{\partial y} \right) 
.
\end{equation}
For instance,
\begin{small}
\begin{equation}
\notag
\frac{\partial \lambda}{\partial y}=\begin{cases}
-\mathrm{Res}_{\lambda}Z^{\text{sys}}_{kk}&	\!\!\!\!\text{shunt-connected $y$ at node-}k\vspace{1.5ex}\\
\left(\!\!\!\begin{array}{c}
\!-\mathrm{Res}_{\lambda}Z^{\text{sys}}_{kk}\!-\!\mathrm{Res}_{\lambda}Z^{\text{sys}}_{ii}\\
\!+\mathrm{Res}_{\lambda}Z^{\text{sys}}_{ki}\!+\!\mathrm{Res}_{\lambda}Z^{\text{sys}}_{ik}
\end{array}\!\!\!\right)& \!\!\!\!\text{branch-connected $y$ as branch-}ki.
\end{cases}
\end{equation}
\end{small}

It can be seen that the admittance participation factor defined in \cite{Zhu2021Participation} is equivalent to the eigenvalue sensitivity of a shunt-connected component, hence can be merged into the $S_\lambda$ and forms a general approach for root-cause analysis. For a component with admittance $y$ in the system, we define the \textit{admittance sensitivity factor} as
\begin{equation}
	s_{\lambda ,y}=-\mathrm{tr}\left( \mathrm{Res}_{\lambda}\mathrm{Z^{\text{sys}}}\cdot \frac{\partial Y^{\text{nodal}}}{\partial y} \right) ^*,
\end{equation}
such that
\begin{equation}
	\Delta\lambda=\langle s_{\lambda ,y},\Delta y(\lambda)\rangle.
\end{equation}

If the sensitivity of $y$ with respect to a parameter of a component $\rho$ is further known, i.e., $\frac{\partial y}{\partial \rho}$, we define the \textit{parameter sensitivity factor} as 
\begin{equation}
\label{rho_sens}
	s_{\lambda ,\rho}=\langle s_{\lambda ,y},\frac{\partial y(\lambda)}{\partial \rho} \rangle
\end{equation}
such that
\begin{equation}
	\Delta \lambda = s_{\lambda ,\rho}\cdot\Delta \rho.
\end{equation}

\begin{figure}[t]
	\centering
	{\includegraphics[width=3.6in]{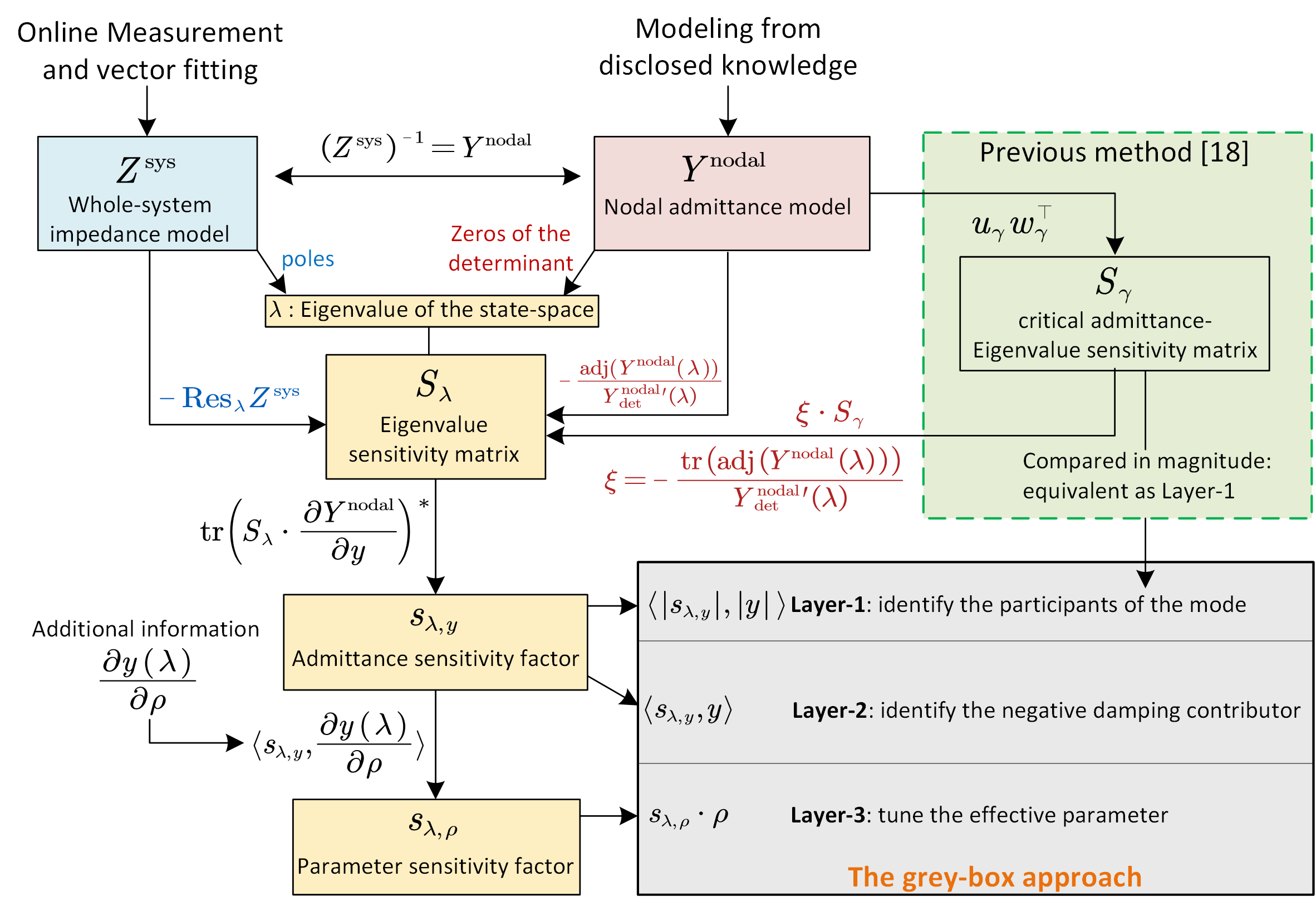}}
	\caption{Structure of the theoretical elements of impedance-based sensitivity analysis and the application of the grey-box approach.}
	\label{fig_instruction}	
\end{figure}

The three-layer grey-box approach can then be extended to the eigenvalue sensitivity analysis, simply by replacing the participation factors with the newly defined sensitivity factors. Such an extension solves the difficulties of analyzing the effect of branch-connected components. The difficulty of comparing complex numbers in $2\times2$ matrix blocks are also solved in the grey-box approach. With the extended grey-box approach, system operators will be able to analyze the origins and the propagation of oscillatory modes. With additional knowledge of $\frac{\partial y}{\partial \rho}$, the system operator can also determine how to tune the parameters with the highest sensitivities to move $\lambda$ leftward in the complex plane to improve the damping of a mode of the system. Since the grey-box approach was described in \cite{Zhu2021Participation}, it is not repeated here. The theoretical elements and the application of the grey-box approach are summarised in \figref{fig_instruction}, which offers a step-by-step guidance on how to extract the sensitivity information from impedance models, either using measurement data or using disclosed models. It should be noted that the methods in this paper consider small-signal conditions, so all of the results above remain true within a small range around steady-state and for the frequency point $s=\lambda$. In a case where the system needs to be tuned over a large range, the methods need to be applied iteratively.

\section{Case Studies}
Three case studies of different scale have been performed. A simple 3-node passive circuit is used to verify the relationships identified between different formulations of eigenvalue sensitivity. A modified IEEE 14-bus test system is used to explore the effectiveness of the methods in assessing system stability and tuning controllers. A modified NETS-NYPS 68-bus network is used to verify the methods in large-scale systems and to provide insights into inter-area modes. The system data, the codes used to generate the simulation results, and all numerical results are available at: \url{https://github.com/Future-Power-Networks/Publications}\cite{SimplusPublication}.
\subsection{Three-node passive circuit}
\begin{figure}
	\centering
	{\includegraphics[width=3.3in]{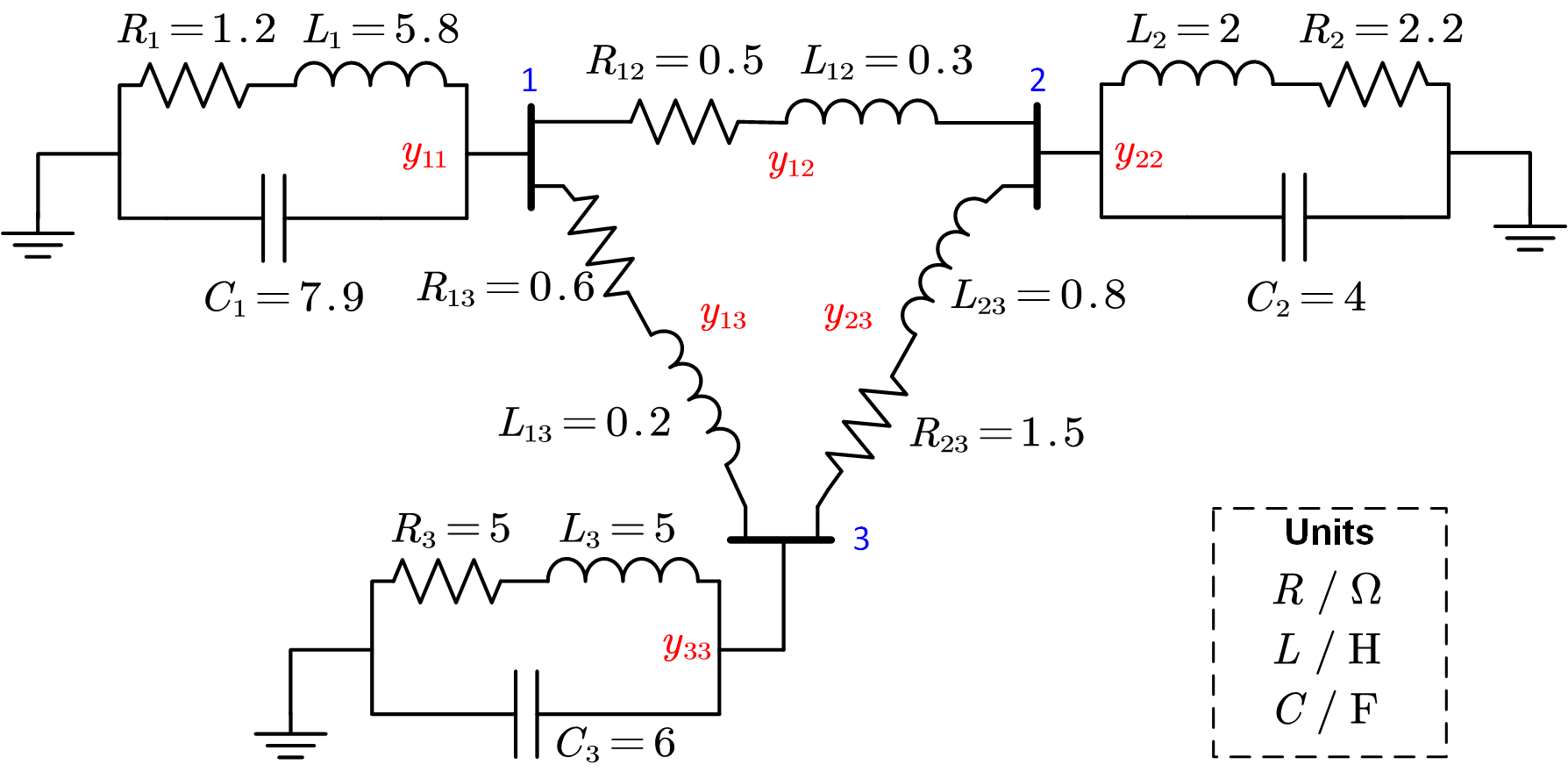}}
	\caption{3-node simple passive circuit.}
	\label{simple_circuit}	
\end{figure}
A single-phase 3-node passive circuit, as shown in \figref{simple_circuit}, is established to demonstrate the advantages of eigenvalue sensitivity analysis in the networked impedance models and to draw a comparison with critical admittance-eigenvalue sensitivity. The nodal admittance model $Y^{\text{nodal}}$ is established as
\begin{equation}
Y^{\text{nodal}}=\left[ \begin{matrix}
y_{11}\!\!+\!\!y_{12}\!\!+\!\!y_{13}&		-y_{12}\!&		-y_{13}\\
-y_{12}\!&		y_{22}\!\!+\!\!y_{12}\!\!+\!\!y_{23}\!&		-y_{23}\\
-y_{13}&		-y_{23}&		y_{33}\!\!+\!\!y_{23}\!\!+\!\!y_{13}\\
\end{matrix} \right],
\end{equation}
where $y_{ik}$ is ane admittance of the component in the system, and \textit{R, L, C} are parameters of the components, such as 
\begin{equation}
\label{RLC_y_represent}
y_{ik}=\begin{cases}
(R_i+sL_i)^{-1}+sC_i \;\;&i=k\\
(R_{ik}+sL_{ik})^{-1} \;\;&i\ne k\\
\end{cases}.
\end{equation}
By calculating the zeros of $Y^{\text{sys}}_{\text{det}}$, we identify 9 eigenvalues for the whole system of which 6 form complex conjugate pairs and all are in the left-half plane. Since the complex eigenvalues represent oscillatory modes, we choose the three pairs of complex conjugate complex eigenvalues for further analysis. For each selected eigenvalue, a corresponding coefficient $\xi$ can be calculated from (\ref{xi_cal}), which defines the relationship between critical admittance-eigenvalue sensitivity and eigenvalue sensitivity.
Because of the repeated information in a conjugate pair, we only consider the values at the upper-half plane, i.e.,
\begin{equation}
\begin{split}
\lambda _1\!&=\!-0.837+ \textit{j}0.968\,\, \mathrm{rad}/S,  \;\;\;  \xi_{1}\!=\!-0.109+\textit{j}0.094
\\
\lambda _2\!&=\!-0.945+ \textit{j}0.270\,\, \mathrm{rad}/S,  \;\;\;  \xi_{2}\!=\!-0.167+\textit{j}0.096
\\
\lambda _3\!&=\!-0.130+\textit{j}0.045\,\, \mathrm{rad}/S,  \;\;\;  \xi_{3}\!=\!-0.082+\textit{j}0.192.
\end{split}
\end{equation}

For each of the three eigenvalues, we identify the two components which have the largest affect on the mode characteristic as noted int the first column of \tabref{Table-1}. The third column shows the critical admittance-eigenvalue sensitivity in magnitude form (Layer-1 of the grey-box), $|\frac{\partial\gamma}{\partial y}||y|$. The penultimate column is the eigenvalue sensitivity in magnitude form, $\langle|s_{\lambda ,y}|,|y|\rangle$. Comparing these two columns, one can see that there is a fixed scalar relationship between them and they therefore contain the same information and both roughly indicate the participation of the components in the modes. However, examining the vector form of the quantities, $\frac{\partial\gamma}{\partial y}$ has a different ratio between the real-part and imaginary-part of the sensitivity than $s_{\lambda,y}$, in other words a different angle. This illustrates the conclusions in \ref{section2.2} that the critical impedance eigenvalue sensitivity does not give an indication of how change in component value will affect damping and frequency of a mode. The result in the final column, $\langle s_{\lambda,y},y \rangle$ (layer-2 of the grey-box) does give guidance on how to stabilize the system by tuning the admittance by scaling-up or scaling down the the value (aligned to its original direction). For instance, by increasing $y_{12}$ proportionally (scaling up), $\lambda_{1}$ will shift to the upper-left direction such that both the damping and natural frequency will increase. 
\begin{table}
\caption{critical admittance-eigenvalue sensitivity and eigenvalue sensitivity for selected components in the grey-box layer 1, 2}
\label{Table-1}
\centering
\setlength{\tabcolsep}{0.8mm}{ \scriptsize
\begin{tabular}{>{\centering}p{1.2cm} |cc|ccc}
\hline
\multirow{2}{*}{\begin{tabular}[c]{@{}c@{}}Mode, \\ component\end{tabular}} & \multicolumn{2}{c|}{Critical admittance-eigenvalue}                            & \multicolumn{3}{c}{State-space eigenvalue}                                                             \\ \cline{2-6} 
& $\frac{\partial\gamma}{\partial y}$ & $|\frac{\partial\gamma}{\partial y}||y|$ & $s_{\lambda ,y}$ & $\langle|s_{\lambda ,y}|,|y|\rangle$ & $\langle s_{\lambda,y},y\rangle$ \\ \hline
$\lambda_{1}$,~$y_{12}$                                                      & 1.747+\textit{j}0.101                        & 4.576                                    & -0.201-\textit{j}0.152    & 0.659                                & -0.036+\textit{j}0.658                   \\
$\lambda_{1}$,~$y_{2}$                                                       & 0.848-\textit{j}0.044                        & 3.969                                    & -0.089-\textit{j}0.084    & 0.571                                & -0.001-\textit{j}0.571                    \\
$\lambda_{2}$,~$y_{13}$                                                      & 1.6981-\textit{j}0.102                       & 4.103                                    & -0.273-\textit{j}0.179    & 0.788                                & -0.597+\textit{j}0.515                    \\
$\lambda_{2}$,~$y_{3}$                                                       & 0.679-\textit{j}0.026                        & 3.802                                    & -0.111-\textit{j}0.069    & 0.730                                & 0.548-\textit{j}0.483                     \\
$\lambda_{3}$,~$y_{1}$                                                       & 0.256+\textit{j}0.053                        & 0.232                                    & -0.031-\textit{j}0.045    & 0.048                                & 0.008+\textit{j}0.048                     \\
$\lambda_{3}$,~$y_{3}$                                                       & 0.442-\textit{j}0.053                        & 0.270                                    & -0.026-\textit{j}0.089    & 0.056                                & -0.009-\textit{j}0.056                    \\ \hline
\end{tabular}
}
\end{table}

From \eqref{RLC_y_represent}, further information of $\frac{\partial y}{\partial \rho}$, where $\rho$ is a parameter (\textit{R, L, C}), can be computed then the layer-3 of the grey-box can be used as illustrated in \tabref{Table-2}. Results are shown for some selected parameters which are influential on $\lambda$. Parameter sensitivity factors $s_{\lambda,\rho}$ are calculated from (\ref{rho_sens}), and a predicted change in eigenvalue $\Delta \lambda_{pr}$ is calculated and shown in third column for a 5\% increment of a parameter based on
\begin{equation}
\Delta \lambda_{\text{pr}} = s_{\lambda,\rho} \cdot \rho \cdot 5\%.
\end{equation}
For comparison, the eigenvalues of the system are re-computed for 5\% increment in parameter and actual the change in value from the original condition is shown in the fourth column as $\Delta \lambda$. The error between the predicted and actual values, in the firth column, were calculated as 
\begin{equation}
\text{error}=\frac{|\Delta \lambda_{\text{pr}} - \Delta \lambda|}{|\Delta \lambda_{\text{pr}}|}.
\end{equation} 

It is clear from \tabref{Table-2} that the whole-system sensitivity analysis provides a useful prediction of the changes of  eigenvalues by tuning specific parameters. The predictions are not perfect because the impedance model is based on linearized small-signal model (linearized around the steady state operating point) and therefore will not be fully accurate for substantial changes of parameter but when the perturbation is small, the error will as well be small. It can be seen that under a 5\% perturbation, the errors are within 20\%, and all changes are in the correct direction. With such predictions available, a system operator can choose the most effective parameters to increase or decrease in a small range, in order to move eigenvalues in the desired direction and adjust either damping or natural frequency or both. If the parameters need to be adjusted over a large range, the grey-box approach would need to be applied iteratively over the path of parameter variations. The case of $\langle|s_{\lambda 3}|,|y_{3}|\rangle$ illustrates an important further point. The value seen in \tabref{Table-1} is of reasonable magnitude but when looking at the effect of individual parameters in the last line of \tabref{Table-2} we observe that it is not possible to change the damping (real part) of $\lambda_{3}$ by adjusting $C_3$, proving the conclusion that Layer-1 can only roughly identify the means to re-tune a mode.

\begin{table}
	\caption{Parameter sensitivity in the grey-box layer 3 and tuning results under 5\% increment}
	\label{Table-2}
	\centering
	\setlength{\tabcolsep}{1mm}{
		\begin{tabular}{>{\centering}p{1.4cm} c c c c c } 
			\toprule
			Mode, parameter	&$s_{\lambda,\rho}\cdot\rho$ &Predicted $\Delta \lambda_{\text{pr}}$ &
			Actual $\Delta \lambda$  & Error\\
			\midrule
			$\lambda_{1},R_{12}$ & -0.619-\textit{j}0.598 & -0.031-\textit{j}0.030  & -0.030-\textit{j}0.031  & 3.63\%   \\
			$\lambda_{1},L_{12}$ & 0.658-\textit{j}0.059  & 0.033-\textit{j}0.003  & 0.031-\textit{j}0.004  & 4.95\%   \\
			$\lambda_{1},C_{2}$  & -0.030-\textit{j}0.625  & -0.001-\textit{j}0.031 & -0.002-\textit{j}0.030  & 3.47\%   \\
			$\lambda_{2},R_{13}$ & 0.759-\textit{j}0.851  & 0.038-\textit{j}0.043  & 0.035-\textit{j}0.038  & 10.43\%  \\
			$\lambda_{2},R_{3}$  & -0.340+\textit{j}0.053  & -0.017+\textit{j}0.003 & -0.017+\textit{j}0.001 & 7.64\%   \\
			$\lambda_{2},C_{3}$  & 0.515-\textit{j}0.572  & 0.0260-\textit{j}0.029  & 0.024-\textit{j}0.027  & 5.67\%   \\
			$\lambda_{3},R_{1}$  & -0.105-\textit{j}0.219 & -0.005-\textit{j}0.011 & -0.005-\textit{j}0.013 & 18.42\%  \\
			$\lambda_{3},L_{1}$  & 0.114+\textit{j}0.114  & 0.006+\textit{j}0.006  & 0.005+\textit{j}0.005  & 11.23\%  \\
			$\lambda_{3},C_{3}$  & -0.004-\textit{j}0.076 & 0.000-\textit{j}0.004  & 0.000-\textit{j}0.004  & 3.84\%   \\			 			
			\bottomrule
		\end{tabular}
	}
\end{table}



\subsection{Modified IEEE 14-bus system}

We now demonstrate the use of sensitivity analysis on a three-phase power system with a mixture of inverters and conventional generators. The case study is based on the IEEE 14-bus network \cite{Kios14Bus}, with three  additional IBRs (Type-IV wind farms) connected to buses 11, 12, 13. The detailed parameters and the control of grid-following inverters are showed in Appendix \ref{14busAppendix}

\begin{figure}[t]
	\centering
	{\includegraphics[width=3.5in]{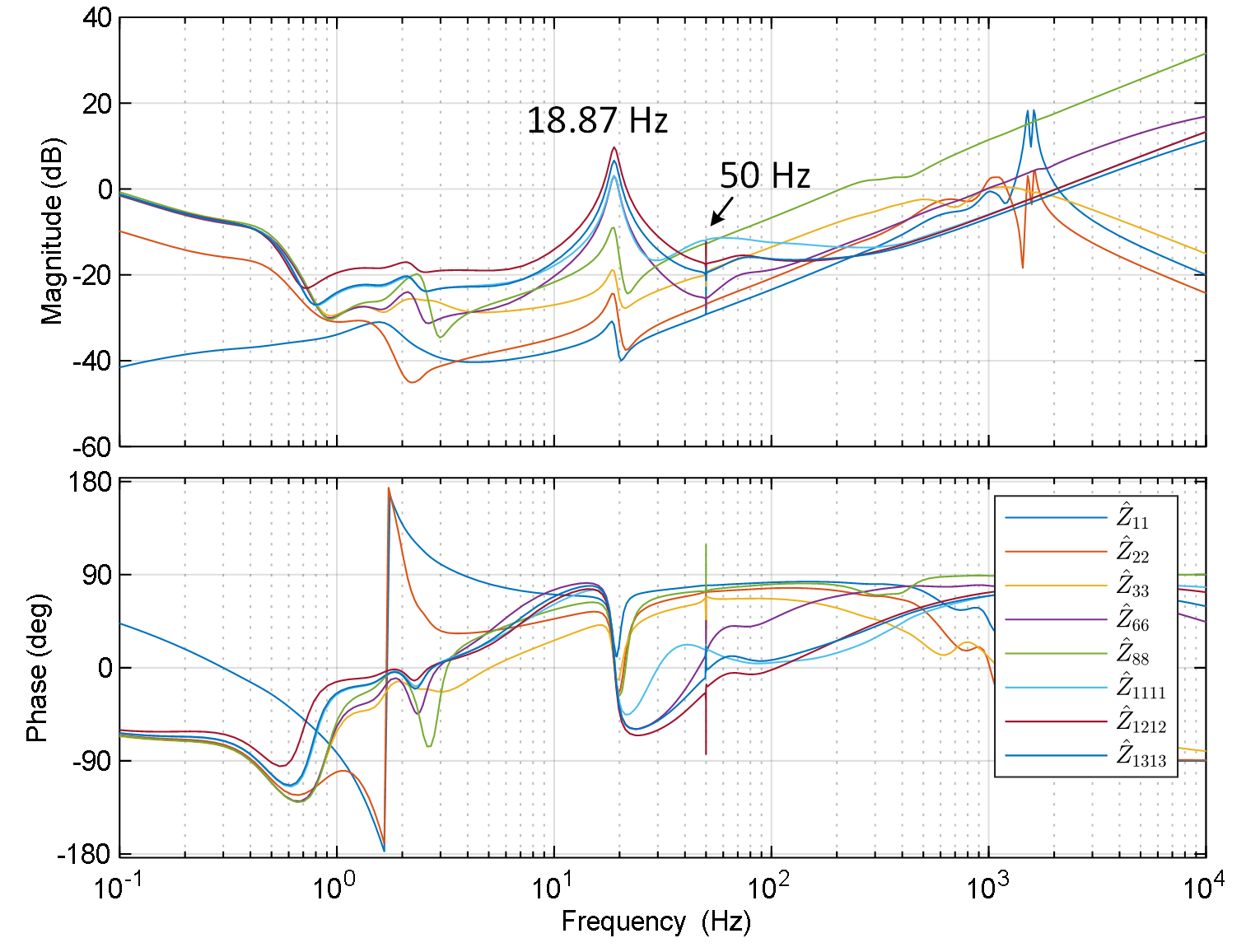}}
	\caption{Bode diagram of whole system impedance $Z^{\text{sys}}_{kk}$ at nodes with sources, presented in $d\text{-}d$ axis.}
	\label{zsys_bode}	
\end{figure}

The whole-system impedance spectra of the network, constructed from the admittance of all apparatus and grid impedances, are displayed in the bode plot in \figref{zsys_bode}. Because this is a three-phase power system modelled in the synchronous $d\text{-}q$ frame, the whole-system admittance $Z^{\text{sys}}_{kk}$ at each node is a $2 \times 2 $ matrix. Only one of the four elements in the matrix (the $d\text{-}d$ term) is displayed since that is sufficient to illustrate the characteristics of the system. Only nodes with sources (SGs or IBRs) present are plotted because the other nodes are passive. A significant peak is observed at 18.87~Hz at all nodes, meaning that the system has an oscillatory mode of 18.87~Hz. We choose this mode for sensitivity analysis. 


\begin{figure}[t]
	\centering
	{\includegraphics[width=3.5in]{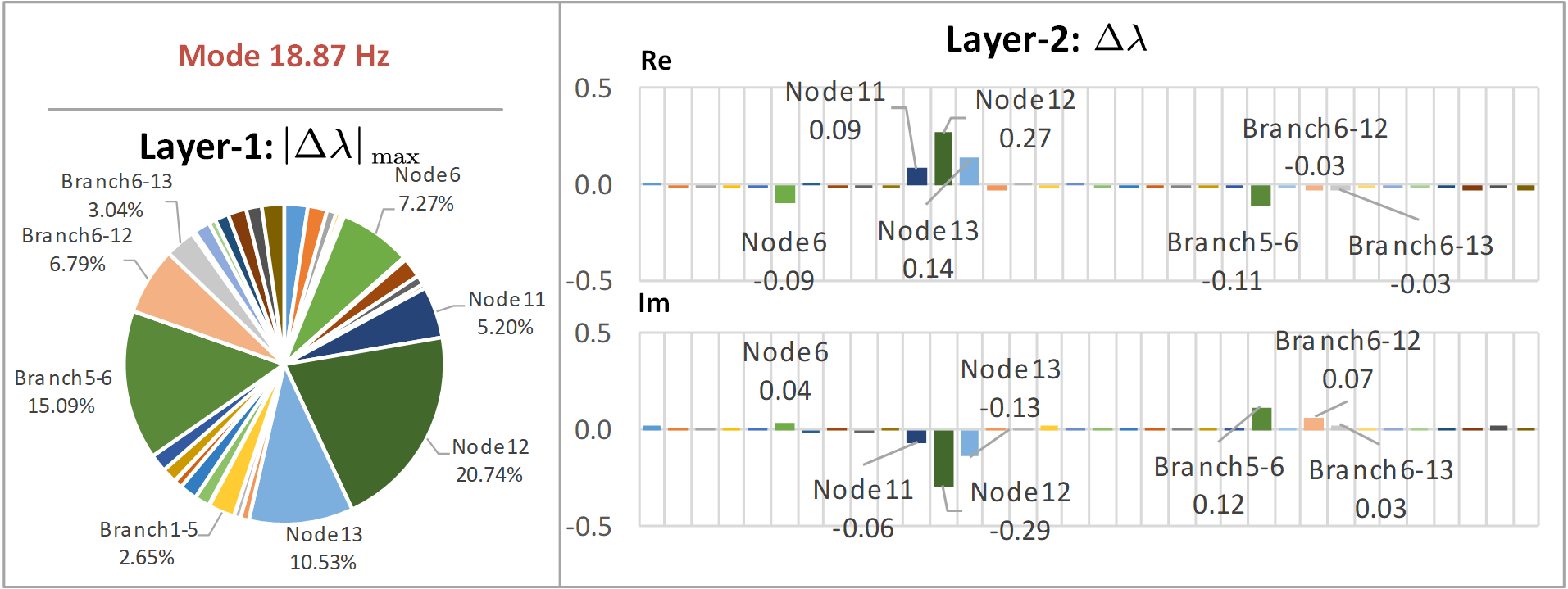}}
	\caption{Grey-box layer-1 and layer-2 results for 18.87~Hz mode, showing that apparatus at nodes 11, 12 and 13, and the adjacent impedances branches have the highest participations. The layer-2 results are normalized to the sum of absolute values.}
	\label{layer12}	
\end{figure}

\figref{layer12} shows the results from the grey-box approach layer-1 and layer-2. It can be seen that nodes 6, 11, 12, 13, branch 5-6, 6-12, 6-13 stand out in the layer-1 pie chart. It indicates that the oscillation is mainly affected by these components. Breakdown of the sensitivity into real and imaginary parts in layer-2 reveals that node 6 and branch 5-6 have negative real parts and node 11, 12, 13 have positive real parts. This indicates that by increasing (scaling up) the admittance of node 6 and branch 5-6, the mode will shift leftwards in the complex plane, i.e., the damping will increase. On the other hand, by scaling up the apparatus admittance at nodes 11, 12, 13, the oscillation will be further exacerbated through reduced damping. Further to the results in layer-1, layer-2 goes on to indicate that for scaling-up of admittance, node 6 and branch 5-6 provide positive damping of the mode, while node 11, 12, 13 provide negative damping. In practical terms, scaling up of the admittance of a windfarm could be achieved by increasing the number of individual turbines operating.

\begin{figure}[t]
	\centering
	{\includegraphics[width=3.5in]{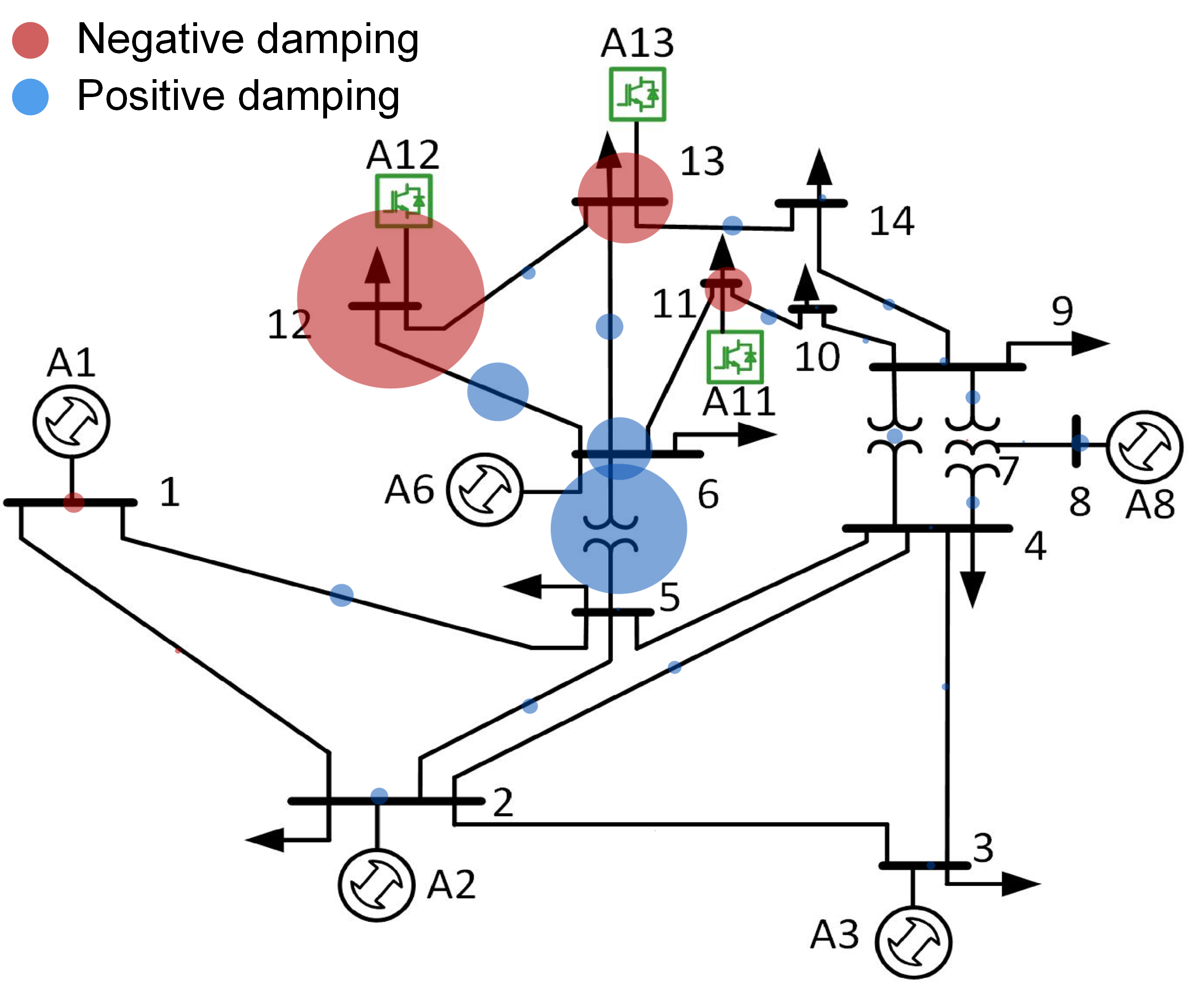}}
	\caption{Modified IEEE 14-bus system with 3 extra IBRs. The circles represent the sensitivity of 18.87~Hz mode to nodes and branches, indicating the origins and the propagation of the oscillation. The circle radii are proportional to the results in the grey-box layer-1, and the filling color is determined by the sign of the real-part in layer-2.}
	\label{propagation}	
\end{figure}

\figref{propagation} maps the influences on the 18.87~Hz mode as either increasing or decreasing damping for an impedance scale-up. It can be seen that damping is reduced by A11, A12 and A13, and increased by A6 and branch 5-6. Further, the main influences and participation in the mode is the upper part of the network with branch 5-6 acting as a boundary.

\begin{figure}[t]
	\centering
	{\includegraphics[width=3in]{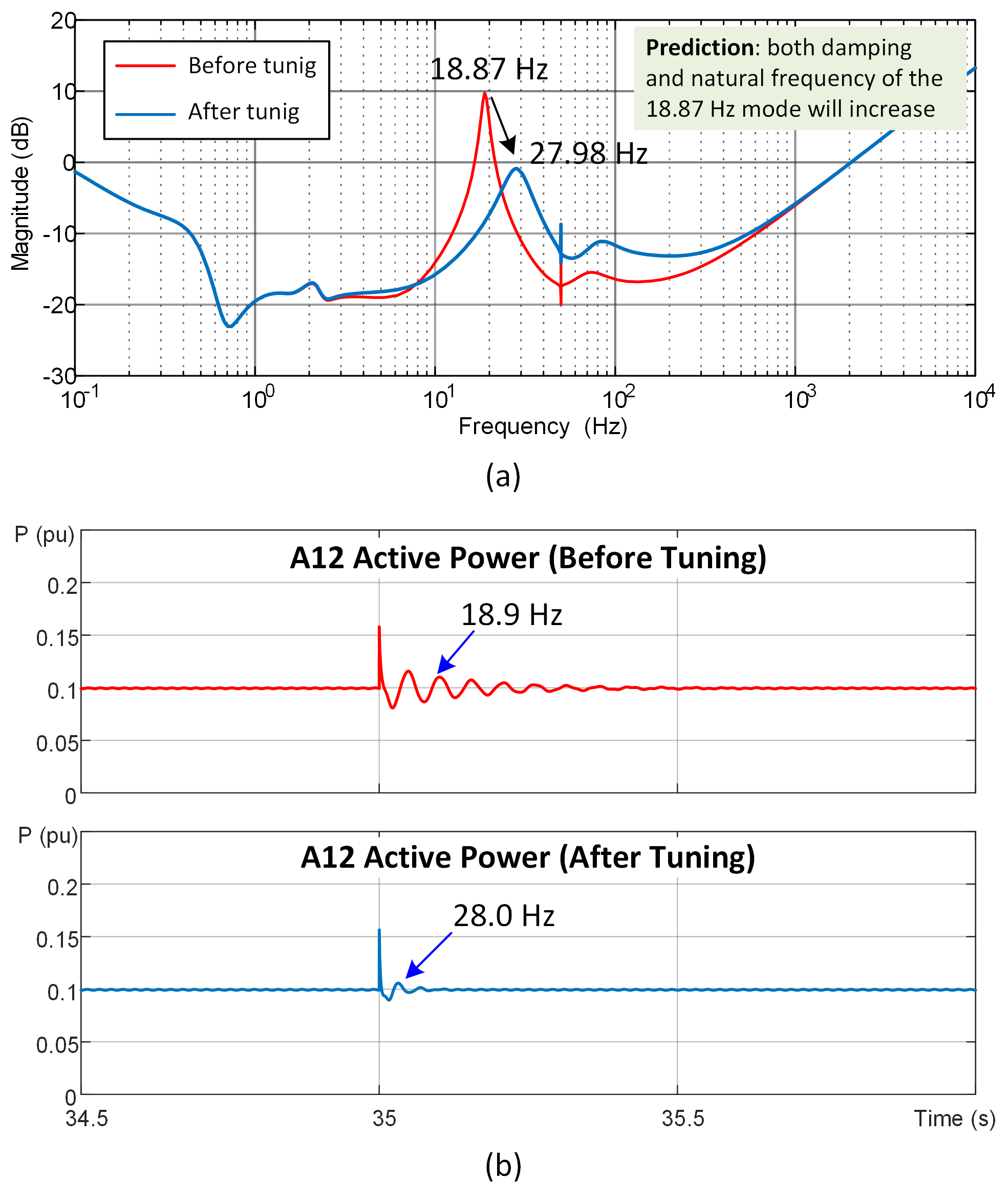}}
	\caption{Re-tuned system with an increase in current control bandwidth of A11, A12 and A13 by 20\%, 50\% and 28.5\%, respectively: (a) bode plot of the $d\text{-}d$ term of $Z^{\text{sys}}_{12,12}$ showing the mode reshaped as predicted; (b) Time domain simulation: active power of A12 during a 100\% demand increase at bus 12 at t=35~s, showing significant improvement in system damping after tuning.}
	\label{layer3-bode}	
\end{figure}

By applying layer-3 of the grey-box, the parameters within A11, A12 and A13 with the greatest influence on the mode can be identified as candidates for tuning. The results are shown in \tabref{Table-3}. It can be seen that the current control bandwidths $f_i$ of the three IBRs are most influential and have a large negative value for the real part of the sensitivity. We can conclude that the low damping of the 18.87~Hz mode is mainly caused by the current control bandwidth of A11, A12 and A13 having been set too low. Layer-3 has narrowed down the root cause of the oscillation to specific parameters in a way that layers 1 and 2 and critical admittance-eigenvalue sensitivity can not. To stabilize the system, we choose to increase $f_i$ of the three IBRs by 20\%, 50\% and 28.5\% for A11, A12 and A13 respectively which will shift the pole leftwards. \figref{layer3-bode} (a) shows the bode plot of $Z^{\text{sys}}_{12,12}$ before and after tuning. It can be seen that the resonance peak is notably flattened but also the frequency is increased from 18.87~Hz to 27.98~Hz as expected from the positive imaginary part of the sensitivities. \figref{layer3-bode} (b) shows the active power output of A12 in a time-domain simulation, in which the load at bus 12 is increased by 100\% at t=35~s which causes a lightly damped oscillation in power flow at 18.9~Hz. It can be clearly seen that in the re-tuned system, the mode is significantly better damped and the oscillation frequency has changed to 28.0~Hz. To tune the mode to a precise characteristic, the whole-system impedance analysis and grey-box sensitivity can be applied iteratively.

\begin{table}[t]
	\caption{Grey-box layer-3: significant parameter sensitivity factors in A11, A12 and A13}
	\label{Table-3}
	\centering
	\begin{tabular}{>{\centering}p{1.4cm} >{\centering}p{1.4cm} c} 
		\toprule
		Apparatus, parameter $\rho$     & original value & $s_{\lambda,\rho} \cdot \rho$\\ 
		\midrule
		A11, $X$       	& 0.03~pu                       & -1.014+\textit{j}1.346     \\
		A11, $f_i$      & 400~Hz                       & -1.818+\textit{j}2.922     \\
		A12, $f_{vdc}$ 	& 10~Hz                        & 1.420+\textit{j}1.465       \\
		A12, $X$        & 0.03~pu                       & -3.171+\textit{j}6.089     \\
		A12, $f_{pll}$  & 10~Hz                        & 1.511+\textit{j}1.118      \\
		A12, $f_i$    	& 300~Hz                       & -5.039+\textit{j}13.317    \\
		A13, $X$     	& 0.03~pu                       & -1.946+\textit{j}2.873     \\
		A13, $f_i$      & 350~Hz                       & -3.379+\textit{j}6.279     \\
		\bottomrule
			\end{tabular}
		\setlength{\tabcolsep}{0.8mm}{
		\begin{tabular}{l p{1.7cm}}
				 & \\
				 & \\
				$X$		& series output reactance\\
				$f_i$	& current control bandwidth\\
				$f_{vdc}$	& dc-link voltage control bandwidth\\
				$f_{pll}$	& phase-lock loop control bandwidth
			\end{tabular}}

\end{table}

\subsection{Modified NETS-NYPS 68-bus system}
\begin{figure}[t]
	\centering
	{\includegraphics[width=2.5in]{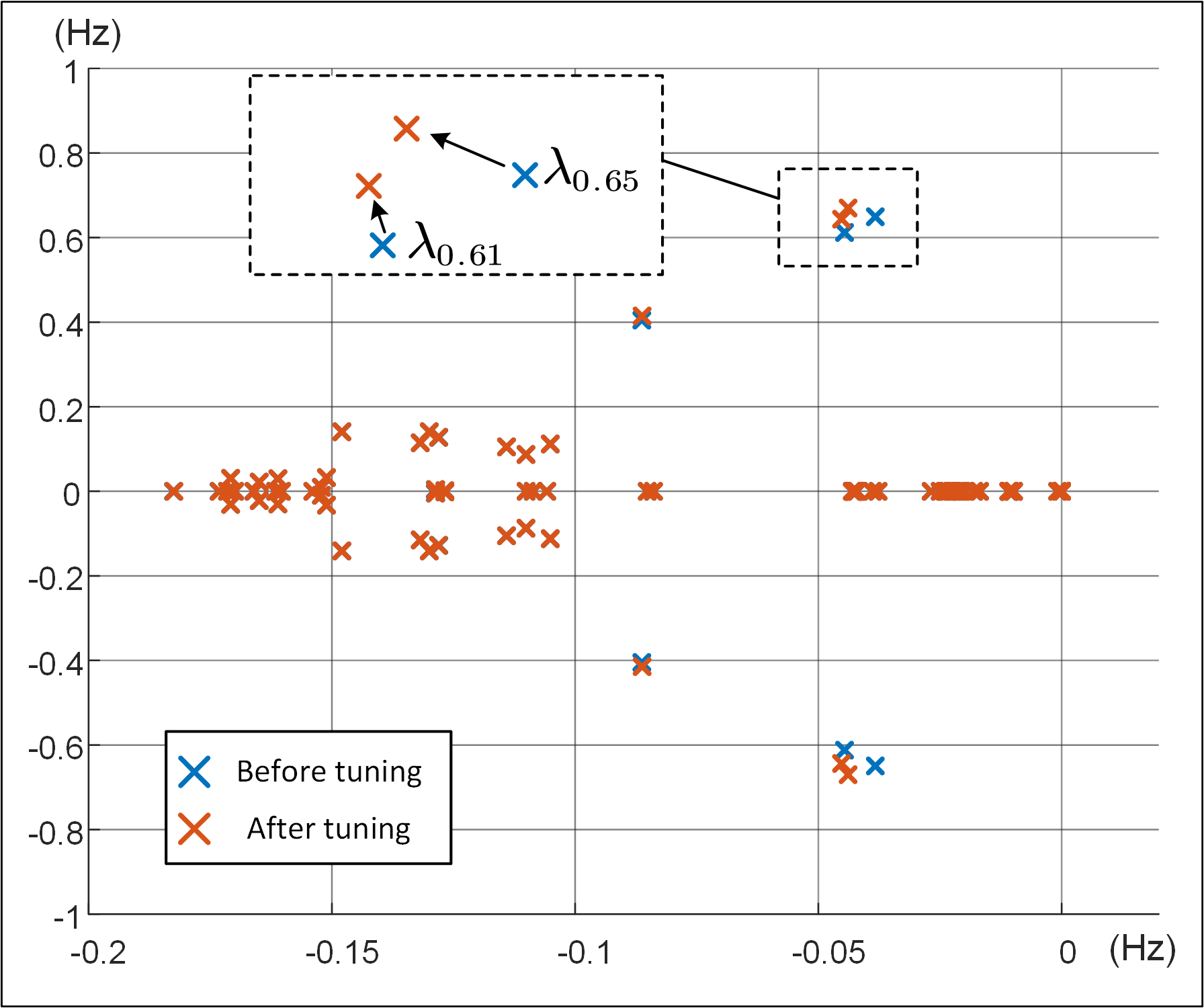}}
	\caption{Eigenvalue plot of the 68-bus system, showing that the inter-area mode is better damped according to the guidance from the grey-box approach.}
	\label{fig_polemap}	
\end{figure}

A modified NETS-NYPS 68-bus system is used here to demonstrate the viability of the proposed theory for large-scale power systems and to prove the ability to analyse inter-area modes. The system structure and all parameters are the same as the system studied in \cite{Zhu2021Participation}, with 15 synchronous generators, 1 grid-forming inverter and 6 grid-following inverters. Compared with the participation analysis in \cite{Zhu2021Participation}, this paper has extended the scope of the sensitivity analysis to include the branches as well as the apparatus at nodes. By calculating the poles of $Z^{\text{sys}}$, eigenvalues of the system can be identified from which a 0.65~Hz mode, $\lambda_{0.65}$, and a 0.61~Hz mode, $\lambda_{0.61}$, are identified as inter-area modes \cite{Singh2013task}. Using the sensitivity analysis and the grey-box approach, the information in \tabref{Table-68} was complied which forms guidance on tuning parameters to improve the damping of these inter-area modes. The analysis reveals that the two inter-area modes are significantly affected by apparatus A1, A13 and A14, and in particular are sensitive to the inertia of the machines. For example, by reducing the inertia of A13, $\lambda_{0.65}$ will shift leftward while $\lambda_{0.61}$ will shift rightward slightly. On the other hand, a reduction in inertia at A14 will improve the damping of $\lambda_{0.61}$ and but worsens that of $\lambda_{0.65}$. The lower part of \tabref{Table-68} shows the predicted change in the modes, $\Delta \lambda$, for a 10\% reduction of inertia of both A13 and A14. The actual change, found by analysing the re-tuned system, is seen to be close to the prediction but not exact because of the linearisation of a non-linear system. The new positions of the modes are confirmed by the pole map in \figref{fig_polemap}. The results confirm the ability of the grey-box sensitivity to produce useful insights into tuning in large systems. However, in the cases of inter-area modes, the sensitivities show that tuning an apparatus parameter involves compromises that lead to only modest improvements in damping. It is known that inter-area modes call for a wide area control solution probably involving  flexible AC transmission system (FACTS) devices \cite{Chaudhuri2009Widearea} but that is outside the scope of this paper.

\begin{table}[]\centering
\caption{68-bus system tuning guidance and results for inter-area modes}
\label{Table-68}
\begin{threeparttable}
\begin{tabular}{cccc}
\toprule
parameter $\rho$ & $s_{\lambda_{0.65},\rho}\cdot \rho$~(Hz)   &      $s_{\lambda_{0.61},\rho}\cdot \rho$~(Hz)         & guidance on $\rho$                   \\ \hline
A1, $H$ \tnote{*}    & -0.0265-\textit{j}0.1386 & 0.0140-\textit{j}0.0100  & $\text{-}$ \\
A13, $H$    & 0.0590-\textit{j}0.1206  & -0.0216-\textit{j}0.1441 & reduce 10\%                     \\
A14, $H$    & -0.0163-\textit{j}0.0617 & 0.0316-\textit{j}0.1504  & reduce 10\%                     \\ \hline
\multicolumn{4}{c}{\emph{Tuning following the guidance}}  \\
& Predicted & Actual & Error \\ \hline
$\Delta \lambda_{0.65}$ &-0.0043+\textit{j}0.0182         &-0.0056+\textit{j}0.0207    & 15.1\%     \\
$\Delta \lambda_{0.61}$ &-0.0010+\textit{j}0.0295         &-0.0006+\textit{j}0.0316    & 7.2\%     \\
\bottomrule
\end{tabular}
\begin{tablenotes}
\footnotesize
\item[*]$H$: the inertial of synchronous generator.
\end{tablenotes}
\end{threeparttable}
\end{table}

\section{Conclusions}
This paper has formalised the relationship between two forms of impedance models and two forms of impedance-based root-cause analysis methods of power systems, and for the first time calculates the value of eigenvalue sensitivity in impedance models. It has shown that the so-called critical admittance-eigenvalue sensitivity in nodal-loop model can indicate which component impedances are most influential on modes of the system but that direction information in the sensitivity is without meaning, whereas eigenvalue sensitivity analysis of whole-system model based on residues of the poles provides meaningful direction information that can be used to assess impact on damping and natural frequency separately. By extension through the grey-box approach, sensitivities to individual equipment parameter can be established which opens up root-cause analysis and parameter turning equivalent to methods available for state-space models. The grey-box approach has been extended to provide sensitivity analysis of series branch components, not just shunt apparatus at nodes, thus facilitating tuning of series compensators. For completeness, a method of calculating the missing complex scaling factor that restores directional meaning to critical admittance-eigenvalue analysis has been established. The unification of approaches to eigenvalue sensitivity in impedance models and the identification of methods for parameter tuning means that tools can be developed for tuning of power systems modes for systems that include IBRs where only black-box impedance models are available.

\appendices
\section{Mathematical Proofs}
\subsection{Proof of critical admittance-eigenvalue $\gamma$}
\label{critcial_proof}
$\gamma=0$ is the only zero eigenvalue of $Y^{\text{nodal}}$, and the determinant $Y^{\text{nodal}}_{\det}(s,\rho)$ can be expressed as the product of all the eigenvalues of $Y^{\text{nodal}}$ at $(\lambda_0, \rho_0)$:
\begin{equation}
    Y^{\text{nodal}}_{\det}(\lambda _0,\rho _0)=\prod_N{\mathrm{eig}\left( Y^{\text{nodal}}(\lambda_0, \rho_0) \right)}=K\cdot \gamma =0,
\end{equation}
where the subscription $0$ means the value before perturbation, $K$ is the product of all none-zero eigenvalues of $Y^{\text{nodal}}$, so that $K$ is a none-zero coefficient. It can be clearly seen that $Y^{\text{nodal}}_{\det}(s,\rho)$ is analytical around $(\lambda_0, \rho_0)$. Now we consider the case where a small perturbation is added on $\rho_0$ and the frequency point $s$ keeps the same, the variation on $Y^{\text{nodal}}_{\det}$ is
\begin{equation}
\label{Delta_gamma}
\begin{split}
      \Delta Y^{\text{nodal}}_{\det}&=\frac{\partial Y^{\text{nodal}}_{\det}}{\partial \rho}\bigg|_{\lambda_0,\rho_0}^{}\Delta \rho 
    \\
    &=K\Delta \gamma +\gamma \Delta K+\Delta \gamma \Delta K\approx K\Delta \gamma   
\end{split}
\end{equation}

Under the same perturbation, the mode $\lambda_0$ is moved to $\lambda_0 + \Delta \lambda$. At the new steady point, the critical admittance-eigenvalue is still zero, such that
\begin{equation}
\label{gamma_new_point}
    Y^{\text{nodal}}_{\det}(\lambda_0+\Delta\lambda, \rho_0+\Delta\rho) = 0
\end{equation}
Applying Taylor's expansion to (\ref{gamma_new_point}), and suppressing the higher orders term gives
\begin{equation}
\label{gamma_expension}
    \frac{\partial Y^{\text{nodal}}_{\det}}{\partial s}\bigg|_{\lambda_0,\rho_0}\Delta \lambda +\frac{\partial Y^{\text{nodal}}_{\det}}{\partial \rho}\bigg|_{\lambda_0,\rho_0}\Delta \rho =0.
\end{equation}

Substituting (\ref{Delta_gamma}) into (\ref{gamma_expension}) leads to
\begin{equation}
\label{gamma_vs_lambda}
\begin{split}
        \Delta \gamma =-\frac{1}{K}\frac{\partial Y^{\text{nodal}}_{\det}}{\partial s}\bigg|_{\lambda_0,\rho_0} \Delta \lambda.\\
        |\Delta \gamma| =\left| -\frac{1}{K}\frac{\partial Y^{\text{nodal}}_{\det}}{\partial s}\bigg|_{\lambda_0,\rho_0}\right |\cdot \left |\Delta \lambda \right|.
\end{split}
\end{equation}
Equation (\ref{gamma_vs_lambda}) proves that when there's a small variation on $\rho$, $\left|\Delta \gamma\right|$ is proportional to $\left|\Delta \lambda\right|$, so that $\Delta \gamma$ could reflect how parameters will affect the mode $\lambda$.

\subsection{Proof of equation (\ref{Sensitivity_matrix})} \label{proof-1}
Expanding $Y^{\text{nodal}}_{\det}(\lambda)$ along row $k$ yields
\begin{equation}
	\label{DetExp}
	Y^{\text{nodal}}_{\det}\left( \lambda \right) =\sum_{i=1}^n{Y^{\text{nodal}}_{ki}C_{ki}},
\end{equation}
where $C_{ki}$ is the cofactor of $Y^{\text{nodal}}_{ki}$. According to (\ref{DetExp}) it is clear to have
\begin{equation}
	\label{partial_Y}
	\frac{\partial Y^{\text{nodal}}_{\det}\left( \lambda \right)}{\partial Y^{\text{nodal}}_{ki}}=C_{ki}.
\end{equation}
Considering a small perturbation is added on a system parameter $\rho$, leading to a perturbation on $Y^{\text{nodal}}_{ki}$ and $\lambda$, i.e.
\begin{equation}
	\begin{split}
		Y^{\text{nodal}}_{ki}&=Y^{\text{nodal}}_{ki0}+\Delta Y^{\text{nodal}}_{ki}
		\\
		\lambda &=\lambda _0+\Delta \lambda,
	\end{split}
\end{equation}
where the subscript $0$ refers to the value before perturbation. At the new steady state, the eigenvalue $\lambda$ still satisfies (\ref{det_zero}), hence we have
\begin{equation}
	\label{AfterPert}
	Y^{\text{nodal}}_{\det}\left( \lambda ,Y^{\text{nodal}}_{ki} \right) =Y^{\text{nodal}}_{\det}\left( \lambda _0+\Delta \lambda ,Y^{\text{nodal}}_{ki0}+\Delta Y^{\text{nodal}}_{ki} \right) =0.
\end{equation}
Since $Y^{\text{nodal}}_{\det}$ is analytical around its zero $\lambda$, applying the first-order Taylor expansion to (\ref{AfterPert}) and suppressing the high-order of infinitesimal items yields
\begin{equation}
	\label{Tayler}
	\frac{\partial Y^{\text{nodal}}_{\det}\left( \lambda \right)}{\partial Y^{\text{nodal}}_{ki}}\Delta Y^{\text{nodal}}_{ki}+Y^{\text{nodal}}_{\det}{}'\left( \lambda \right) \Delta \lambda =0,
\end{equation}
where 
\begin{equation}
\label{Y_det_prime}
	Y^{\text{nodal}}_{\det}{}'\left( \lambda \right) \triangleq \frac{dY^{\text{nodal}}_{\det}\left( s \right)}{ds}\bigg|_{s=\lambda}.
\end{equation}
Substituting (\ref{partial_Y}) into (\ref{Tayler}) yields the result of $\frac{\partial \lambda}{\partial Y_{ki}}$
\begin{equation}
	\frac{\Delta \lambda}{\Delta Y^{\text{nodal}}_{ki}}=\frac{\partial \lambda}{\partial Y^{\text{nodal}}_{ki}}=S_{\lambda ,ik}=-\frac{C_{ki}}{Y^{\text{nodal}}_{\det}{}'\left( \lambda \right)},
\end{equation}
where $\mathrm{C}$ is the cofactor matrix of $Y^{\text{nodal}}(\lambda)$ and $C_{ki}$ is its element. $S_{\lambda}$ can then be deduced as
\begin{equation}
S_{\lambda}=-\frac{1}{Y^{\text{nodal}}_{\det}{}'\left( \lambda \right)}\mathrm{C}^\top =-\frac{1}{Y^{\text{nodal}}_{\det}{}'\left( \lambda \right)}\text{adj}(Y^{\text{nodal}}(\lambda)).
\end{equation}

\subsection{Proof of equation (\ref{adj_Y})}	\label{proof-2}
Since $\lambda$ is considered as a non-repeated eigenvalue of the system, the rank of $Y^{\text{nodal}}(\lambda)$ is $N-1$, hence $Y^{\text{nodal}}(\lambda)$ has one and only one zero-eigenvalue $\gamma$. Accordingly, the rank of its adjugate matrix $\text{adj}(Y^{\text{nodal}}(\lambda))$ is 1, with only one non-zero-eigenvalue $\gamma^\dagger$. It is known that rank-1 matrix can be expressed as the outer product of two vectors, such that
\begin{equation}
\label{b_0}
	\text{adj}( Y^{\text{nodal}}(\lambda))=x\otimes y=xy^\top,
\end{equation}
where $x$ and $y$ are two column-vectors of $N$-order. Now we prove $ u_{\gamma}$, which is the right eigenvector of $Y^{\text{nodal}}(\lambda)$ corresponding to $\gamma$, is proportional to $x$. 

For $\gamma^\dagger$ we have
\begin{equation}
\label{b_1}
    	\text{adj}( Y^{\text{nodal}}(\lambda))\cdot  u_{\gamma}^\dagger=\gamma^\dagger  \cdot  u_{\gamma}^\dagger,
\end{equation}
where $ u_{\gamma}^\dagger $ is a non-zero right-eigenvector corresponding to $\gamma^\dagger$. Left-multiplying $Y^{\text{nodal}}$ in (\ref{b_1}) and rearranging the equation yields
\begin{equation}
\label{b_2}
    Y^{\text{nodal}}(\lambda)\cdot  u_{\gamma}^\dagger = 0\cdot  u_{\gamma}^\dagger,
\end{equation}
where we use fact $Y^{\text{nodal}}(\lambda)\cdot\text{adj}(Y^{\text{nodal}}(\lambda))=Y_{\det}(\lambda)=0$. (\ref{b_2}) proves that $ u_{\gamma}^\dagger$ is also a right-eigenvector of $Y^{\text{nodal}}(\lambda)$ corresponding to $\gamma$, i.e., the non-zero eigenvectors $ u_{\gamma}^\dagger$ and $ u_{\gamma}$ are linear combinations of each other. Reversely, $ u_{\gamma}$ is also the right eigenvector of $\text{adj}(Y^{\text{nodal}}(\lambda))$ corresponding to $\gamma^\dagger$. Combined with (\ref{b_0}) it is clear to have 
\begin{equation}
\label{b_3}
\begin{split}
        \text{adj}( Y^{\text{nodal}}(\lambda))\cdot  u_{\gamma} &= xy^\top  u_{\gamma} = \gamma^\dagger  \cdot  u_{\gamma}\\
        x &= \frac{\gamma^\dagger }{y^\top  u_{\gamma}} \cdot  u_{\gamma}.
\end{split}
\end{equation}
Since $\frac{\gamma^\dagger }{y^\top  u_{\gamma}}$ is a scalar, $x$ is proportional to $ u_{\gamma}$. Similarly, we can prove $y^\top$ is proportional to $ w_{\gamma}^\top$. As a result,
\begin{equation}
\label{b_4}
    xy^\top=\eta \cdot  u_{\gamma}  w_{\gamma}^\top,
\end{equation}
where $\eta$ is a scalar. From (\ref{b_0}) it is clear that
\begin{equation}
\label{b_5}
    \text{tr}\left( \text{adj}\left( Y^{\text{nodal}}\left( \lambda \right) \right) \right) = y^\top x = \eta \cdot  w_{\gamma}^\top u_{\gamma}.
\end{equation}
Because $ u_{\gamma}$ and $ w_{\gamma}$ are normalized as $ w_{\gamma}^\top u_{\gamma}=1$, $\eta=\text{tr}(\text{adj}(Y^{\text{nodal}}(\lambda)))$. Substituting $\eta$ into (\ref{b_4}) yields equation (\ref{adj_Y}).

\section{Modified IEEE 14-bus system parameters} \label{14busAppendix}
\begin{figure}[t]
	\centering
	{\includegraphics[width=3in]{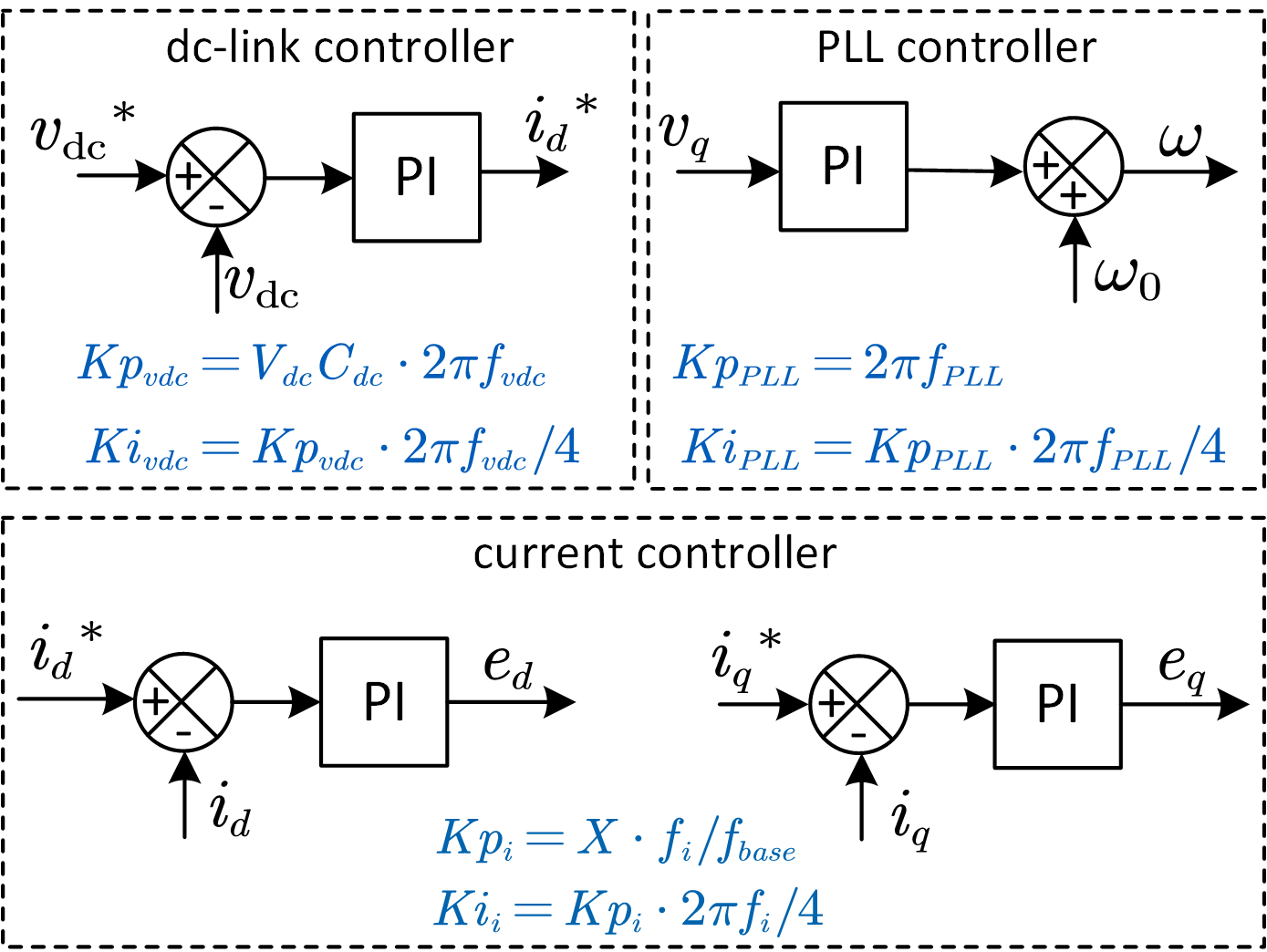}}
	\caption{Control diagram of the grid-following inverter, $e_{dq}$ is the inverter leg voltage, $v_{dq}$ is the filter terminal voltage.}
	\label{fig_VSI}
\end{figure}

\setcounter{table}{0}
\renewcommand{\thetable}{B\arabic{table}}
This appendix gives the detailed parameters of the modified 14-bus system studied in this paper. The parameters of the synchronous generators are from the dynamic model built by KIOS centre at University of Cyprus \cite{Kios14Bus}. The parameters of the three extra grid-following inverters are given in \tabref{Table-B1}, and the control diagrams and PI parameters are shown in \figref{fig_VSI}.
\begin{table} 
\centering
\caption{Parameter of grid-following inverters}
\begin{tabular}{llll}
\toprule
\multirow{2}{*}{Parameters} & \multicolumn{3}{c}{Values} \\
                            & A11     & A12     & A13    \\ \hline
$V_{dc}$, dc-link voltage                         & 2.5~pu     & 2.5~pu     & 2.5~pu    \\
$C_{dc}$, dc-link capacitor                         & 1.25~pu    & 1.25~pu    & 1.25~pu   \\
$X$, series output reactance                           & 0.03~pu    & 0.03~pu    & 0.03~pu   \\
$R$ series output resistance                           & 0.01~pu    & 0.01~pu    & 0.01~pu   \\
$f_{vdc}$, dc-link control bandwidth                      & 10~Hz      & 10~Hz      & 10~Hz     \\
$f_{PLL}$, PLL control bandwidth                      & 10~Hz      & 10~Hz      & 10~Hz     \\
$f_{i}$, current control bandwidth                      & 400~Hz     & 300~Hz     & 350~Hz    \\ \bottomrule

\end{tabular}
\label{Table-B1}
\end{table}

\ifCLASSOPTIONcaptionsoff
  \newpage
\fi

\bibliographystyle{IEEEtran}
\bibliography{References_PES}
\begin{IEEEbiography}[{\includegraphics[width=1in,height=1.25in,clip,keepaspectratio]{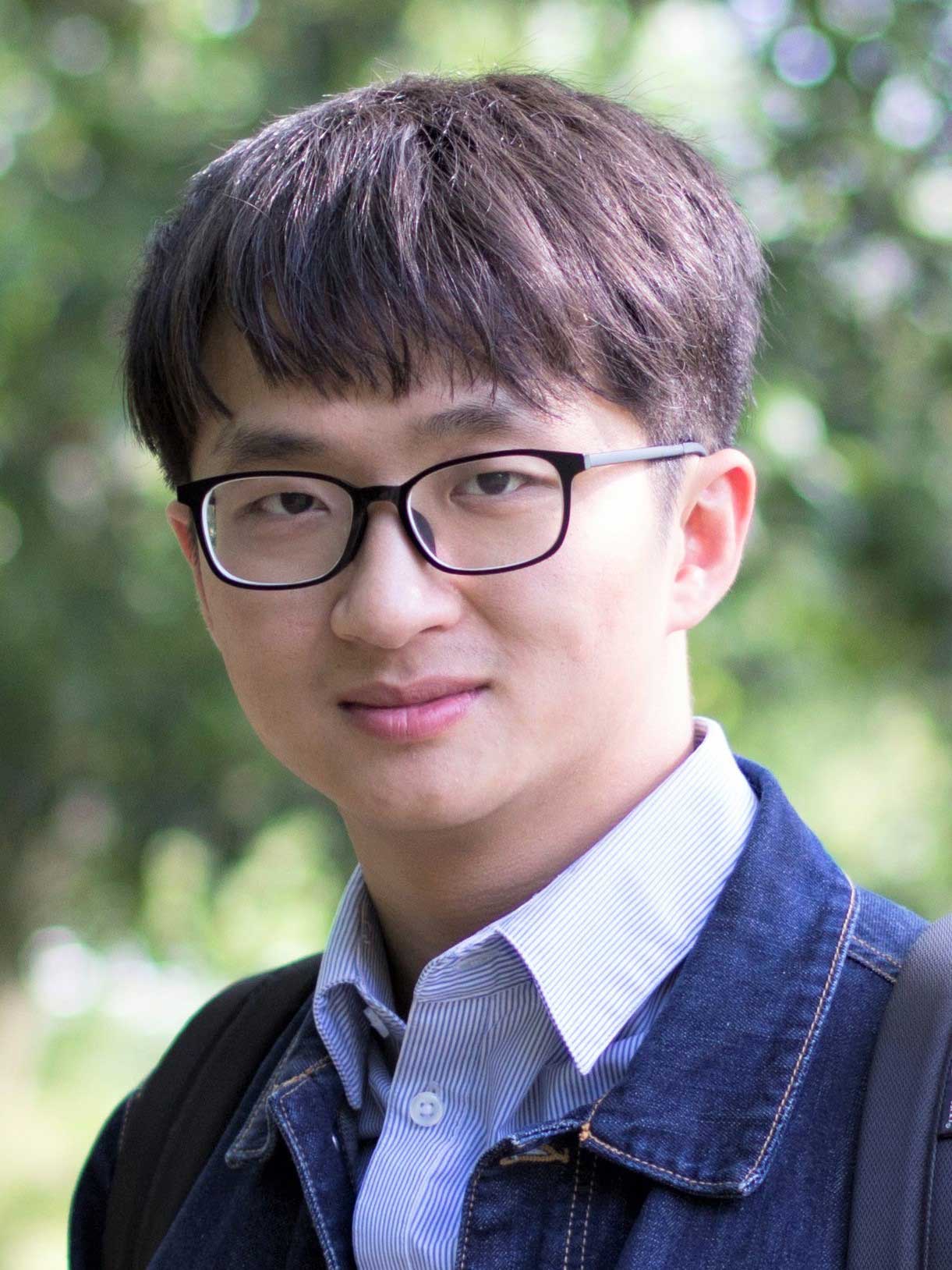}}]{Yue Zhu}
(S'21) received the B.Eng and M.Sc degrees in electrical engineering, Zhejiang University, Hangzhou, China, in 2016 and 2019 respectively. He is currently a PhD student and a research assistant at Department of Electrical and Electronic Engineering, Imperial College London, UK. He is also a visiting student at University of Bath, UK. His present research focuses on impedance-based stability analysis of power systems, and the noise evaluation for impedance measurement.
 \end{IEEEbiography}

\begin{IEEEbiography}[{\includegraphics[width=1in,height=1.25in,clip,keepaspectratio]{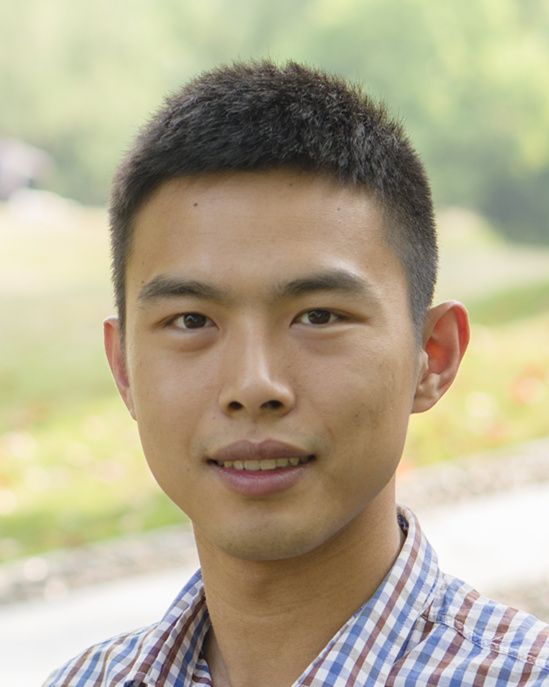}}]{Yunjie Gu}
(M'18-SM'20) received the B.Sc. and the Ph.D. degree in Electrical Engineering from Zhejiang University, Hangzhou, China, in 2010 and 2015 respectively. He was a Consulting Engineer at General Electric Global Research Centre, Shanghai, from 2015 to 2016. After that, he joined Imperial College London as a Research Associate and was an EPSRC-funded Innovation Fellow (award EP/S000909/1) from 2018 to 2020. He is now a Lecturer at University of Bath, and an Honorary Lecturer at Imperial College.
His research interests include the fundamental theories and computational tools for dynamic analysis of power-electronic-based renewable power systems, and new technologies for stability enhancement.
\end{IEEEbiography}

\begin{IEEEbiography}[{\includegraphics[width=1in,height=1.25in,clip,keepaspectratio]{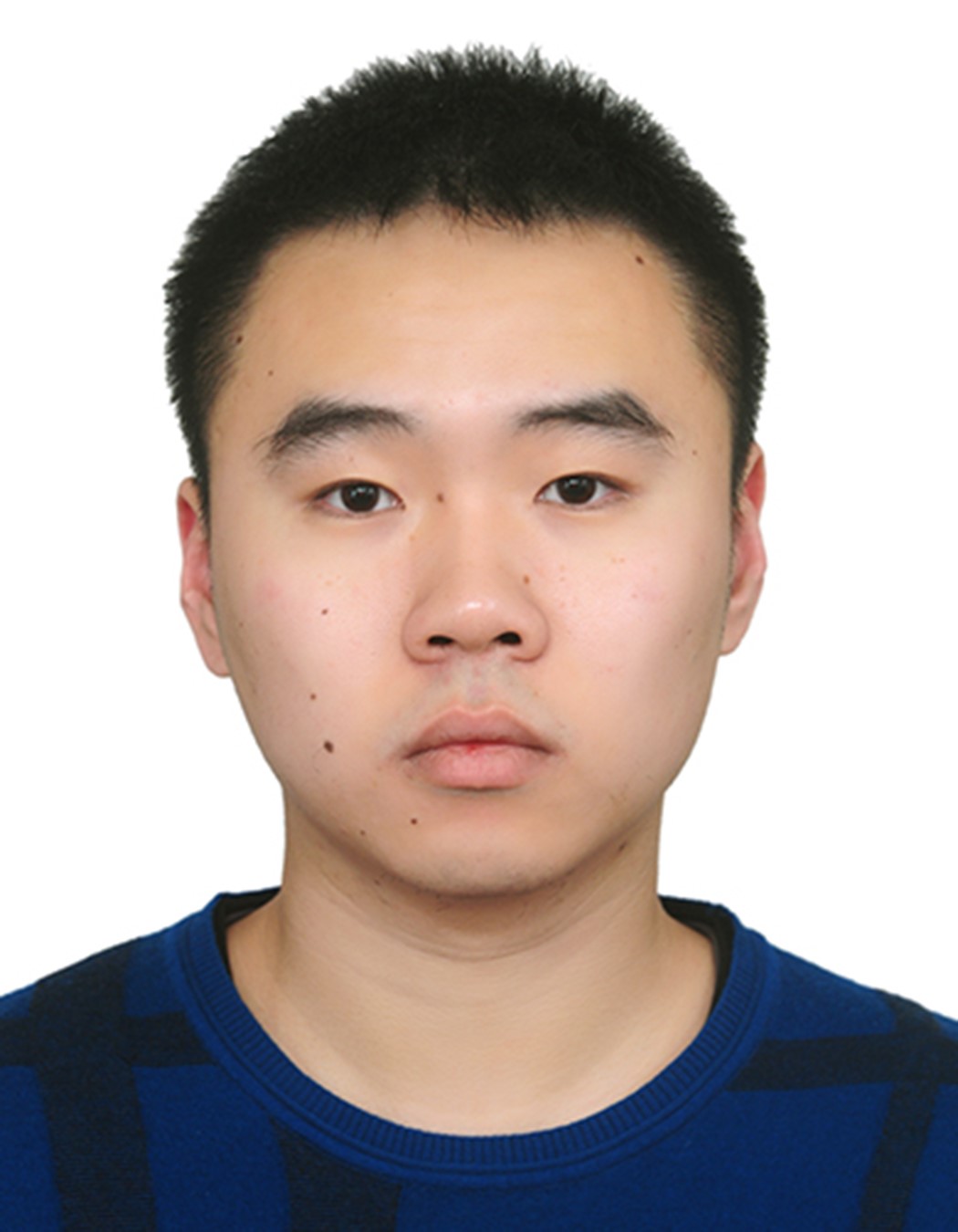}}]{Yitong Li}
(S'17-M'21) received the B.Eng degrees in electrical engineering from Huazhong University of Science and Technology, China, and the University of Birmingham, UK, in 2015. He received the M.Sc degree in future power networks and the Ph.D. degree in electrical engineering from Imperial College London, UK, in 2016 and 2021 respectively. He is currently a Research Associate at Imperial College London, UK. He is also a Visiting Researcher at University of Bath, UK. His current research interests include control of power electronic converters and analysis of power system dynamics.
\end{IEEEbiography}
\begin{IEEEbiography}[{\includegraphics[width=1in,height=1.25in,clip,keepaspectratio]{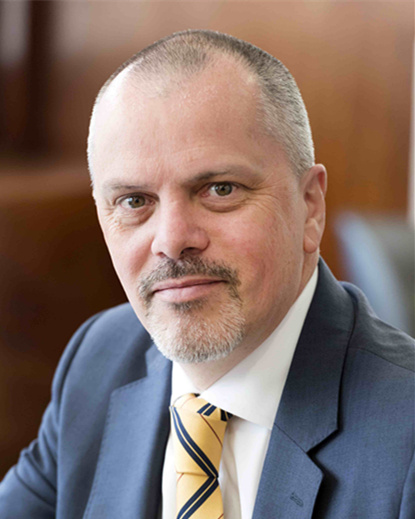}}]{Timothy C. Green}
(M'89-SM'02-F'19) received a B.Sc. (Eng) (first class honours) from Imperial College London, UK in 1986 and a Ph.D. from Heriot-Watt University, Edinburgh, UK in 1990. He is a Professor of Electrical Power Engineering at Imperial College London, and Co-Director of the Energy Futures Lab with a role fostering interdisciplinary energy research. His research interest is in using the flexibility of power electronics to accommodate new generation patterns and new forms of load, such as EV charging, as part of the emerging smart grid. In HVDC he has contributed converter designs that reduce losses while also providing control functions assist AC system integration. In distribution systems, he has pioneered the use of soft open points and the study of stability of grid connected inverters. Prof. Green is a Chartered Engineering the UK and a Fellow of the Royal Academy of Engineering.
\end{IEEEbiography}

\end{document}